\pdfoutput=1
\documentclass[11pt]{article}

\usepackage[final]{acl}
\usepackage{natbib}
\usepackage{times}
\usepackage{latexsym}
\usepackage{amsmath}
\usepackage[T1]{fontenc}
\usepackage[utf8]{inputenc}

\usepackage{microtype}

\usepackage{inconsolata}

\usepackage{graphicx}

\usepackage{subcaption}
\usepackage{cite}
\usepackage{amsmath,amssymb,amsfonts}
\usepackage{algorithmic}
\usepackage{graphicx}
\usepackage{textcomp}
\usepackage{multicol}
\usepackage{multirow}
\usepackage{xcolor}
\usepackage{pifont}

\usepackage{booktabs, multirow} % for borders and merged ranges
\usepackage{soul}% for underlines
\usepackage{changepage,threeparttable} % for wide tables
\usepackage{amsmath,amssymb,amsfonts}
\usepackage{textcomp}
\usepackage{xcolor}
\usepackage{amsmath,tabularx}
\usepackage{makecell}
\usepackage{tablefootnote}
\usepackage{changepage,threeparttable}
\usepackage{hyperref}
\usepackage{times}

\usepackage{hyperref}
\usepackage{url}

\title{InSerter: Speech Instruction Following with Unsupervised Interleaved Pre-training}

\author{%
  Dingdong Wang$^1$\thanks{~~Equal contribution.},  Jin Xu$^2$\footnotemark[1], Ruihang Chu$^1$, Zhifang Guo, Xiong Wang, Jincenzi Wu$^1$ \\ \textbf{Dongchao Yang$^1$, Shengpeng Ji, Junyang Lin$^2$}
  \\
  $^1$The Chinese University of Hong Kong, $^2$Alibaba Group \\
  \texttt{dingdongwang@link.cuhk.edu.hk} 
}

\newcommand{\method}{InSerter}

\begin{document}
\maketitle
\begin{abstract}

Recent advancements in speech large language models (SpeechLLMs) have attracted considerable attention. Nonetheless, current methods exhibit suboptimal performance in adhering to speech instructions. Notably, the intelligence of models significantly diminishes when processing speech-form input as compared to direct text-form input. Prior work has attempted to mitigate this semantic inconsistency between speech and text representations through techniques such as representation and behavior alignment, which involve the meticulous design of data pairs during the post-training phase. In this paper, we introduce a simple and scalable training method called \textbf{InSerter}, which stands for \textbf{In}terleaved \textbf{S}p\textbf{e}ech-Text Rep\textbf{r}esen\textbf{t}ation Pr\textbf{e}-t\textbf{r}aining. InSerter is designed to pre-train large-scale unsupervised speech-text sequences, where the speech is synthesized from randomly selected segments of an extensive text corpus using text-to-speech conversion. Consequently, the model acquires the ability to generate textual continuations corresponding to the provided speech segments, obviating the need for intensive data design endeavors. To systematically evaluate speech instruction-following capabilities, we introduce SpeechInstructBench, the first comprehensive benchmark specifically designed for speech-oriented instruction-following tasks. Our proposed model \method~achieves SOTA performance in SpeechInstructBench and demonstrates superior or competitive results across diverse speech processing tasks. SpeechInstructBench is available at \url{https://huggingface.co/datasets/ddwang2000/SpeechInstructBench}.

\end{abstract}

\section{Introduction}

\begin{figure}[!t]
    \centering
    \includegraphics[width=1\linewidth]{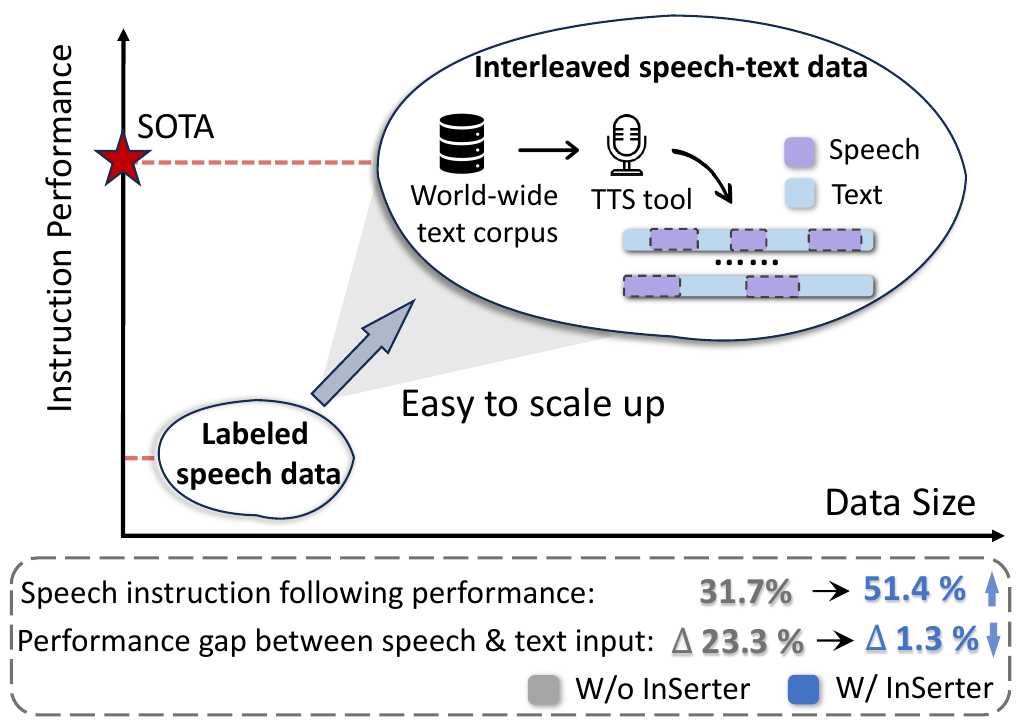}
    % \caption{InSerter, a scalable interleaved pre-training framework, improves speech instruction-following accuracy from 31.7\% to 51.4\%, reduces the accuracy gap between speech-form and text-form inputs from 23.3\% to 1.3\% compared to non-interleaved baseline.}
    \caption{InSerter leverages large-scale interleaved data for pre-training. It improves instruction-following accuracy (SpeechInstructBench closed-ended standard version) given speech-form input from 31.7\% to 51.4\%, further narrowing the gap with text-form inputs from 23.3\% to 1.3\% compared to non-interleaved baseline.}
    \label{fig:teaser}
    \vspace{-2mm}
\end{figure}

Recently, speech large language models~(SpeechLLMs) have garnered significant attention, as they hold promise for enabling more natural and intuitive human-machine interactions~\citep{ji2024wavchat, Qwen-Audio,Qwen2-Audio,kyutai2024moshi, tang2024salmonn}.  SpeechLLMs receive speech inputs and directly follow speech-based instructions to perform responses. However, due to the intrinsic differences between speech and text representations, it is challenging for models to effectively comprehend the consistency in semantic content expression across different modalities. This discrepancy leads to difficulties in accurately following spoken instructions compared to text-based ones. 
% For example, given the same instruction ``Write a 100-word story with key word “Dream” used 3 times.'', a model might correctly follow this in text form but struggle with spoken variations.

Enabling the speech modality to inherit the cognitive capabilities of text represents a fundamental challenge in the domain of end-to-end speech instruction following. Prior research has attempted to address this challenge through two approaches: representation alignment~\citep{held2024diva} and behavior alignment~\citep{wang2024blsp, held2024diva}. However, representation alignment encounters limitations stemming from the inherent differences in granularity, such as the varying sequence lengths and the discrete nature of text tokens compared to continuous speech frames. Consequently, the direct alignment of these distinct representations often results in the degradation of critical original speech features, including tone, energy, and pitch. On the other hand, behavior alignment focuses on training models to learn consistent response sequences or distributions given either speech input or its transcripts during the post-training phase. Nonetheless, these approaches face significant scalability challenges due to their selected training stage, the complexity of data construction, and overall training efficiency. Therefore, there remains substantial scope for further improvement.

The emergence of intelligence in text is fundamentally based on unsupervised next-token prediction during the pre-training phase. Specifically, the training objective is to predict the subsequent continuation token based on a sequence of preceding text. Therefore, the motivation for our work is to enable speech sequences to inherit the intelligence characteristic of text by requiring the model to execute the following task: \textit{given a speech sequence, predict the next text token or segment continuation during the pre-training phase}. By interleaving speech and text sequences in training samples, we allow for multi-turn text-speech alignment and thus boost training efficiency.

To this end, we propose \textbf{InSerter}, which stands for \textbf{In}terleaved \textbf{S}p\textbf{e}ech-Text Rep\textbf{r}esen\textbf{t}ation Pr\textbf{e}-t\textbf{r}aining. InSerter employs text-to-speech to synthesize speech segments from randomly selected portions of large-scale text pre-training corpora, thereby ensuring scalability. InSerter provides several key advantages: 1) it facilitates dynamic interactions between speech and text modalities through an interleaved training format; 2) it enhances training efficiency by enabling simultaneous learning of multiple segment relationships during pre-training; and 3) it offers high scalability by leveraging existing text corpora via text-to-speech conversion.

To comprehensively evaluate models' performance on speech instruction following, we introduce SpeechInstructBench. This benchmark is characterized by the incorporation of 1) linguistic variations in speech, which encompass spontaneous speech phenomena such as disfluencies and self-corrections, as well as diverse accents; 2) acoustic variations that include speaker-specific traits such as pitch modulation, speech rate, emotional tone, and energy levels; and 3) environmental variations, which account for background noise to simulate real-world interaction conditions. Our experiments show that InSerter achieves state-of-the-art performance on SpeechInstructBench and significantly improves results across multiple benchmarks within VoiceBench~\citep{chen2024voicebench}.

% , achieving improvements of 0.49/5 and 0.20/5 on AlpacaEval~\citep{alpaca_eval} and CommonEval, respectively alongside substantial increases of 27.66\%, 23.55\%, 27.69\%, and 0.99\%  score increases on SD-QA~\citep{faisal-etal-21-sdqa}, MMSU, OpenBookQA, and AdvBench~\citep{robey2021adversarial} tasks,

We summarize our contributions as follows:
% \begin{itemize}
%     \item We propose InSerter, a conceptually simple, powerful, and scalable pre-training algorithm to significantly improve speech instruction following capacilities of SpeechLLMs.
%     \item We present SpeechInstructBench, the first comprehensive and bilingual (English-Chinese) benchmark designed to evaluate speech instruction-following abilities.
%     \item InSerter attains state-of-the-art performance on SpeechInstructBench and achieves significant improvement across various speech processing tasks in VoiceBench~\citep{chen2024voicebench}.
% \end{itemize}

\noindent $\bullet$ We propose InSerter, a conceptually simple, powerful, and scalable pre-training algorithm to significantly improve speech instruction following capacilities of SpeechLLMs.

\noindent $\bullet$ We present SpeechInstructBench, the first comprehensive and bilingual (English-Chinese) benchmark designed to evaluate speech instruction-following abilities.

\noindent $\bullet$ InSerter attains state-of-the-art performance on SpeechInstructBench and achieves significant improvement across various speech processing tasks in VoiceBench~\citep{chen2024voicebench}.

    % evaluate the robustness of models in following speech commands under complex acoustic variations such as background noise and self-correction. This benchmark encompasses seven distinct subtasks and comprises a total of 5,550 audio questions.

\begin{figure*}[!t]
    \centering
    \includegraphics[width=1\linewidth]{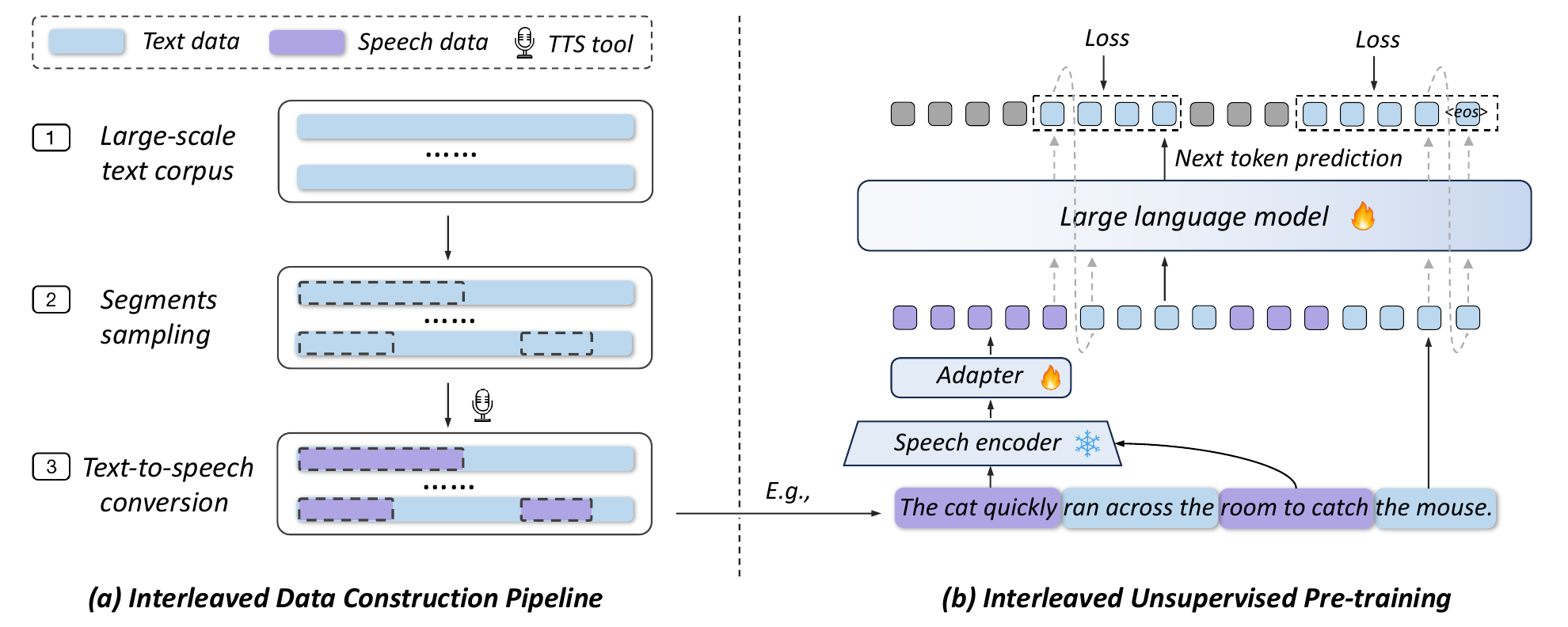}
    % \vspace{-4mm}
    \caption{(a) Data construction pipeline for generating large-scale interleaved speech-text data for pre-training; (b) Interleaved unsupervised pre-training. The gray areas indicate tokens with no loss calculation during training.}
    \label{fig:arch}
    \vspace{-2mm}
\end{figure*}

\section{Related Work}

\paragraph{SpeechLLMs.}
Current SpeechLLMs are generally categorized by their speech representation strategies, using either discrete speech tokens or continuous speech features as input. The first category converts speech inputs into discrete tokens~\citep{kyutai2024moshi, chen2024emova, veluri2024syncllms, zhang2023speechgpt, rubenstein2023audiopalm, borsos2023audiolm, xie2024miniomni2opensourcegpt4ovision}, leveraging self-supervised speech pre-training to expand the vocabulary of LLMs with learned speech representations. Some studies~\citep{kyutai2024moshi, nguyen2024spiritlm, zeng2024scaling} incorporate synthetic interleaved data for pre-training. However, there are two notable distinctions between these works and InSerter. First, the primary objective of these studies is to enhance the quality and end-to-end coherence of speech generation, while InSerter aims to improve models' speech instruction-following capabilities. Second, existing works rely on discrete speech sequence modeling, whereas InSerter is specifically designed for models with continuous speech representations, such as Qwen-Audio series~\citep{Qwen-Audio,Qwen2-Audio}.

The second approach processes continuous speech signals directly~\citep{chen2023lauragpt,gong2023joint,gong2023listen,ma2024embarrassingly,xue2024echatemotion,Qwen-Audio,Qwen2-Audio,tang2024salmonn}. This strategy avoids discretization loss and retains rich speech characteristics, such as intonation, rhythm, and speaker-specific attributes. Typically, a pre-trained speech encoder extracts high-dimensional speech representations, which are then mapped to LLM-compatible embeddings via adapter layers or cross-modal alignment mechanisms.

\paragraph{Improving speech instruction following.}

Previous research has investigated two main approaches to enhance speech instruction following in SpeechLLMs. The first approach is representation alignment~\citep{held2024diva}, which aligns continuous speech representations with their corresponding text embeddings such as L2 loss. 
Yet, this method struggles with the inherent granularity mismatch between discrete text tokens and continuous speech frames, often degrading critical speech features. In contrast, InSerter does not impose any prior constraints on speech representations, facilitating a higher potential for performance and avoiding the loss of information beyond semantic content.

The second approach, behavior alignment~\citep{wang2024blsp, held2024diva}, trains the model to generate identical response sequences or distributions for speech input and its transcripts during post-training. While effective, several limitations persist. First, the reliance on high-quality paired data restricts both scalability and data coverage. Second, the additional self-distillation loss increases training complexity. Third, data efficiency is limited, as each instance can only optimize a single speech sequence's text continuation. Instead, InSerter enhances training efficiency by utilizing large-scale text during the pre-training phase. Moreover, it requires no new loss forms, making the training process simpler and more scalable.

\paragraph{Benchmarks for speech instruction following.}
Evaluating SpeechLLMs-based chat assistants is challenging due to their broad capabilities and the inadequacy of existing benchmarks in measuring human preferences~\citep{yang2024air,chen2024voicebench,adubench2025}. Current benchmarks, including Voicebench~\citep{chen2024voicebench} and ADU-bench~\citep{adubench2025}, primarily evaluate general audio understanding and dialogue capabilities, such as safety, knowledge base, and multilingual support. However, they lack specific frameworks for assessing speech-based instruction following. To bridge this gap, we introduce SpeechInstructBench, a comprehensive benchmark designed to assess models' robustness in following speech instructions under various noisy conditions.

% Several recent benchmarks have been developed to evaluate the general audio understanding and dialogue capabilities. Among them,  Voicebench~\citep{chen2024voicebench} assesses the general abilities of SpeechLLMs as voice assistants, such as safety and general knowledge. Similarly, ADU-bench~\citep{adubench2025} evaluates basic audio dialogue abilities, including different skills and multilingual support. However, they lack specific frameworks for assessing speech-based instruction following. To address this gap, we introduce SpeechInstructBench, the first comprehensive benchmark designed for evaluating the robustness of models in following speech instructions under various settings.

\section{Method}

InSerter aims to improve model performance through large-scale unsupervised pre-training with interleaved speech-text data. To this end, we construct a large-scale interleaved speech-text dataset (Sec.~\ref{subsec:dataset}) and design a novel pre-training strategy to significantly improve the model's instruction following abilities (Sec.~\ref{subsec:training}).

\subsection{Interleaved Data Construction}
\label{subsec:dataset}

To achieve effective interleaved pre-training, we develop a three-stage data construction pipeline shown in Fig.~\ref{fig:arch}(a), the details of each stage are:

\noindent
\textbf{(i) Large-scale text corpus collection and pre-processing.} To collect diverse and high-quality text corpus, we aggregated data from internally collected long-form and dialogue datasets. The original data is pre-processed utilizing strategies including removing special characters and standardization via regular expressions to improve the quality, giving rise to a filtered text corpus comprising around 610 billion text tokens. Detailed statistics of the corpus are provided in Appendix~\ref{subsec:pertraindata}.

\noindent
\textbf{(ii) Segments sampling.} Given the source text corpus, we sample portions of text and convert them into speech segments. Our sampling strategy operates at two granularity levels: word-level and sentence-level. For word-level sampling, we randomly select certain words as the segments for conversion, maintaining a minimum length of five words to preserve semantic coherence. For example, regarding ``[The weather is really nice] today, I want to go for [a walk in the park] in the afternoon'', the portions in brackets represent the speech segments we sampled, while the remaining are text segments. For sentence-level sampling, we use punctuation marks as delimiters to sample sentences as speech segments randomly. 
% An example is ``[The weather is really nice today], I want to go for a walk in the park in the afternoon''. In this way, sentence-level strategy encourages the model to learn longer-range dependencies between text and speech during pre-training. 

% For word-level sampling, we randomly select certain words as the segments for conversion, and we set a minimum length threshold of five words for both segment types to ensure that both the speech and text segments retain their core semantic integrity. For example, regarding ``[The weather is really nice] today, I want to go for [a walk in the park] in the afternoon'', the portions in brackets represent the speech segments we sampled, while the remaining are text segments. For sentence-level sampling, we use punctuation marks as delimiters to sample sentences as speech segments randomly. An example is ``[The weather is really nice today], I want to go for a walk in the park in the afternoon''. In this way, sentence-level strategy encourages the model to learn longer-range dependencies between text and speech during pre-training. Note that the segment sampling ratio $\eta$ will control the proportion of content converted to speech. We set $\eta$ to 30\% for optimal results, as ablated in Sec.~\ref{subsec:ablation}.

\noindent
\textbf{(iii) Text-to-speech conversion.} We convert the sampled speech segments by CosyVoice 2.0 model~\citep{du2024cosyvoice2scalablestreaming} with multi-speaker voice prompts. Specifically, to enhance the diversity of the synthesized speech, we select 10,000 unique voice prompts from an extensive timbre library by evaluating Word Error Rate (WER) and WV-MOS~\citep{Andreev_2023}. The final generated speech corpus consists of 301,540 hours of speech data, together with the remaining text data to form the interleaved speech-text dataset.

\begin{table*}[t]
\resizebox{\textwidth}{!}{%
\begin{tabular}{cccl}
\toprule
\textbf{Types} & \textbf{Tasks} & \textbf{Num} & \multicolumn{1}{c}{\textbf{Instruction Examples}} \\ \hline
\textbf{Closed-Ended} & \begin{tabular}[c]{@{}c@{}}Standard \\ Accent\\ Background\\  Disfluency\\ Paralanguage \end{tabular} & \begin{tabular}[c]{@{}c@{}} 7718 (En) \\ 5208 (Cn)\end{tabular} & \begin{tabular}[c]{@{}l@{}}Include keywords \{keyword\} in your response at least three times. \\ You should not use \{forbidden words\} in the response.\\ Entire output should be wrapped in \{JSON\} format. \\ Your entire response should be in English, \{capital letters only\}. \\ Uh...finish response with um…the exact phrase \{end phrase\}. And … and…no other words should follow this phrase.\\Give two different responses and should be separated by \{6 continuous asterisk symbols\}. \end{tabular} \\ \hline

\textbf{Open-Ended} & \begin{tabular}[c]{@{}c@{}}Content \\ Format\\ Style\\  Linguistic\\ Situation \end{tabular} & \begin{tabular}[c]{@{}c@{}} 500 (En) \\ 500 (Cn)\end{tabular} & \begin{tabular}[c]{@{}l@{}}Recommend me \{5 Chinese films released before 1990\}.\\ Give me suggestions on keeping health, \{use bullet point\} in your answer.\\ How did US states get their names? Please respond in the writing \{style of Shakespeare\}.\\Change the first person to \{the third person\} in the given sentence.\\ \{I have a small child at home\}. How can I increase my productivity while working from home?\end{tabular} \\ \hline

\textbf{Adjustment} & \begin{tabular}[c]{@{}c@{}} Daily Instructions\end{tabular} & \begin{tabular}[c]{@{}c@{}} 250 (En) \\ 250 (Cn)\end{tabular} & \begin{tabular}[c]{@{}l@{}}Write a memoir about campus life. Hmm, change it to a graduation speech.\\ Write a product promotional copy, Ah, hold on, I would like you to write a user guide, with clear steps.\\ Write a funny joke. Oh no, it should be a heartwarming short story. \end{tabular} \\ 

\bottomrule
\end{tabular}%
}
\caption{The statistics and examples of the SpeechInstructBench.}
\label{instruct-example}
\end{table*}

\subsection{Training Strategy}
\label{subsec:training}

The prepared interleaved data is adopted to pre-train our SpeechLLM, which can significantly enhance the model’s ability to follow speech instructions through text-speech alignment. Then we conduct a supervised fine-tuning stage using dialogue data to enhance the model's interactive performance. We detail the two-stage training below.

\noindent
\textbf{Stage 1: Pre-training on interleaved data.}
InSerter guides the model to align speech embeddings with text tokens through next-token prediction. As shown in Fig.~\ref{fig:arch}(b), we process speech segments with a speech encoder and an adapter, resulting in continuous speech representations and interleaved with text segments. 
Let \(S = (S_1, \ldots, S_i, \ldots, S_N)\) denote the sequence of speech embeddings from \(N\) speech segments, and \(Y = (Y_1, \ldots, Y_i, \ldots, Y_N)\) denote the corresponding text token sequences. Each text token segment \(Y_i\) consists of \(M_i\) tokens \((Y_{i,1}, \ldots, Y_{i,M_i})\).
The training objective is to minimize the cross-entropy loss for predicting the next text token conditioned on both the preceding speech embeddings and text tokens:
\begin{equation*}
L_{\text{CE}} = - \sum_{i=1}^{N} \sum_{j=1}^{M_i} \log p(Y_{i,j} \mid S_{1:i}, Y_{<i,j}; \theta),
\end{equation*}
where \(Y_{i,j}\) represents the predicted \(j\)-th text token in \(i\)-th text token segment. Each \(Y_{i,j}\) is predicted conditional on all preceding speech embeddings \(S_{1:i}\) (from the first to the \(i\)-th segment) and all previous text tokens \(Y_{<i,j}\) (up to the \(j\)-th text token in the \(i\)-th text token segment). \(\theta\) is all trainable model parameters, including those within the adapter (Q-Former) and the LLM. Importantly, as indicated in Fig.~\ref{fig:arch}(b), cross-entropy loss is computed only for tokens corresponding to ground-truth text segments, while tokens associated with speech segments are masked out during loss calculation.

In addition to interleaved data, in the pre-training stage, we follow the same multi-task speech data categories~\citep{Qwen2-Audio, Qwen-Audio} and collect the speech data to strengthen the bidirectional speech-text relationships. Text-only corpora are also added to preserve the model's strong text generation quality. We balance these data in a proportion of 40\% (interleaved), 30\% (multi-task speech data), and 30\% (text-only).

\noindent
\textbf{Stage 2: Supervised fine-tuning.} After the pre-training stage aligns speech and text modalities, we conduct supervised fine-tuning (SFT) to enhance the model's dialogue capabilities. We collect data from speech dialogue and text dialogue and empirically incorporate 50\% text data during this stage. The total amount of SFT data is 20K samples.

\section{SpeechInstructBench}
\label{sec:benchmark}

Instruction-following capability refers to a model's ability to precisely interpret and execute natural language directives from users while adhering to all stated constraints and requirements~\citep{zhou2023instructionfollowingevaluationlargelanguage}. However, the current existing benchmarks on SpeechLLMs primarily evaluate response quality rather than instruction following accuracy. To bridge this gap, we present SpeechInstructBench, a comprehensive and bilingual (English-Chinese) speech instruction-following benchmark for SpeechLLMs. As shown in Table~\ref{instruct-example}, this benchmark encompasses three evaluation sub-tasks: closed-ended instructions, open-ended instructions, and adjustment instructions.

\subsection{Closed-Ended Instruction}
Closed-ended instruction questions objectively evaluate SpeechLLM's ability to follow predefined rules using IFEval~\citep{zhou2023instructionfollowingevaluationlargelanguage} metrics: Prompt-level strict-accuracy, Prompt-level loose-accuracy, Instruction-level strict-accuracy, and Instruction-level loose-accuracy. Specifically, we adapt the English IFEval~\citep{zhou2023instructionfollowingevaluationlargelanguage} dataset by selecting 432 questions suited for speech modality. We create corresponding 402 Chinese questions and modify them to accommodate spoken language-specific instructions.

All SpeechInstructBench speech data is generated using the Azure AI Speech system.
To capture the complexity of real-world speech interactions, we synthesize speech data for the closed-ended instruction incorporating the following elements: 1) Accents: we implemente 13 English and 8 Chinese regional accents, such as Indian and Irish variants for English, while Cantonese and Sichuan dialects for Chinese; 2) Environmental Noise: we integrate 80 distinct background sounds (e.g., markets, subway stations, background music) combined with clean speech to simulate the real-world acoustic conditions; 3) Paralinguistic Features: we vary six emotions (happy, sad, terrified, angry, unfriendly, general), multiple timbres (20 in Chinese, 25 in English), and diverse prosodic parameters (speech rate, pitch, energy); 4) Spontaneous Disfluencies: based on the disfluency definitions provided in~\citep{ferreira2004disfluencies}, we utilize GPT-4o with designed prompts to generate corresponding disfluent versions of instructions, incorporating six disfluency types: filled pauses, unfilled pauses, repetitions, self-corrections, false starts, and filter words. The detailed statistics and prompts are provided in Appendix~\ref{subsec:appendspeechinstructbenchdesign}.

\subsection{Open-Ended Instruction} 
For more complex instruction-following tasks that extend beyond deterministic program verification, such as writing "in Shakespeare's style" or "providing examples," we build up an open-ended instruction set. Each element is decomposed into binary sub-questions and could be answered by "YES" for criterion adherence or "NO" for non-compliance. We employ GPT-4o as the evaluator and the evaluation metrics for open-ended instructions are consistent with closed-ended instructions, except with the loose criteria.

For data construction, we collect raw data from previous works~\citep{jiang-etal-2024-followbench, qin2024infobench} and curate 1000 instructions, along with 5,000 decomposed sub-questions, to construct the final closed-ended instruction set. The set covers five categories: 1) content (i.e., explicit restrictions on the response content); 2) format (i.e., response format requirements); 3) style (i.e., response style requirements); 4) linguistic (i.e., dictate the use of particular language structures and terms); and 5) situation (i.e., specific situation/background information added to the question).

\begin{table*}[t]
\centering
\footnotesize
\vspace{-0.2cm}
\renewcommand{\arraystretch}{1.2}  % 使表格行高变大
\setlength{\tabcolsep}{4pt}  % 使列之间的空隙变窄

\resizebox{\textwidth}{!}{  % Scale the table to the text width
\begin{tabular}{ll|c|c|c|c|c|c}
\toprule
\multirow{2}{*}{\textit{SpeechLLMs}} & \multirow{2}{*} & \multicolumn{1}{c}{AlpacaEval} & \multicolumn{1}{|c|}{CommonEval} & \multicolumn{1}{c|}{SD-QA} & \multicolumn{1}{c|}{MMSU} & \multicolumn{1}{c|}{OepnBookQA} & \multicolumn{1}{c}{AdvBench} \\ \cmidrule{3-4} \cmidrule{5-8}

& & \makecell{GPT$\uparrow$ \\ (S / T)} & \makecell{GPT$\uparrow$\\(S / T)} & \makecell{Panda$\uparrow$ $\vert$ GPT$\uparrow$\\(S / T) $\vert$ (S / T)} & \makecell{ACC$\uparrow$\\(S / T)} & \makecell{ACC$\uparrow$\\(S / T)} & \makecell{RR$\uparrow$\\(S / T)}  \\ \midrule

BLSP &  & 3.76 / 4.13 & 3.13 / 4.17 & 46.65 / 62.92 $\vert$ 23.83 / 41.24 & 25.01 / 37.57 & 21.97 / 52.08 & 8.65 / 7.88 \\

GLM-4-Voice &  & 3.97 / 4.57 & 3.42 / 4.10 & 52.44 / 59.49 $\vert$ 36.98 / 38.65 & 39.75 / 42.81 & 53.41 / 61.09 & 88.08 / 85.00 \\

Mini-Omni &  & 1.95 / 2.23 & 2.02 / 2.55 & 23.69 / 26.04 $\vert$ 4.16 / 7.23 & 24.69 / 26.74 & 26.59 / 30.55 & 37.12 / 86.35 \\

Mini-Omni2  &  & 2.32 / 2.65 & 2.18 / 2.86 & 11.03 / 13.02 $\vert$ 7.59 / 9.76 & 24.27 / 27.13 & 26.59 / 32.09 & 57.50 / 92.88 \\

Megrez  &  & 3.50 / 4.23 & 2.95 / 3.90 & 34.53 / 54.61 $\vert$ 25.95 / 53.12 & 27.03 / 49.93 & 28.35 / 57.58 & 87.69 / 96.15 \\

DIVA &  & 3.67 / \textbf{4.68} & 3.54 / \textbf{4.29} & \textbf{62.39} / \textbf{78.30} $\vert$ 51.72 / \textbf{74.50} & 25.76 / 63.31 & 25.49 / 76.70 & \textbf{98.27} / \textbf{99.23} \\

Qwen2-Audio & \multirow{2}{*} & 3.74 / 4.11 & 3.43 / 3.77 & 41.77 / 61.66 $\vert$ 29.66 / 40.69 & 35.72 / 45.02 & 49.45 / 67.91 & 96.73 / 96.73 \\

\textbf{InSerter} & \multirow{2}{*} & \textbf{4.23} / 4.39 & \textbf{3.63} / 4.05 & 59.13 / 61.66 $\vert$ \textbf{57.32} / 64.20 & \textbf{59.27} / \textbf{64.03} & \textbf{77.14} / \textbf{83.52} & 97.69 / 97.50   \\

\bottomrule
\end{tabular}
}  % End resizebox
\caption{The performance of SpeechLLMs on VoiceBench. T and S refer to the model
performance with text-form and speech-form input respectively. For AdvBench, RR stands for Refusal Rate.}
\label{tab:main-vb}
\end{table*}

\subsection{Adjustment Instruction}
In practical applications, users often need to adjust or correct instructions. This scenario involves two key components: the original instruction (termed the erroneous instruction) and its revised version (the modified instruction). To meet this need, we develop an Adjustment Instruction set for the modification or correction of instructions. Using GPT-4o, we generate a set of 500 adjustment instructions and introduce two evaluation metrics: Instruction Adherence Rate (IAR) and Error Correction Rate (ECR). $IAR = \frac{1}{n} \sum_{i=1}^{n} f(resp_i, inst_i)$, where $n$ is the total number of instructions, $resp_i$ is the model's response to the $i$-th instruction, and $inst_i$ is the $i$-th modified instruction. The function $f(resp_i, inst_i)$ equals 1 when the response adheres to the modified instruction, and 0 otherwise. $ECR = \frac{1}{m} \sum_{j=1}^{m} g(resp_j, inst_j)$, where $m$ is the total number of erroneous instructions, $resp_j$ is the model's response to the $j$-th erroneous instruction, and $inst_j$ is the $j$-th erroneous instruction. $g(resp_j, inst_j) = 1$ if the model successfully corrects the erroneous instruction, and $0$ otherwise.

Notably, for both open-ended and adjustment instruction sets, we conduct human evaluations and compare them with GPT-4o judgments, demonstrating consistent agreement with over 92\% accuracy across randomly selected cases. This will ensure the reliability of our evaluation results.

\begin{table*}[t]
\centering
\footnotesize
\vspace{-0.2cm}
\renewcommand{\arraystretch}{1.2}  % 使表格行高变大
\setlength{\tabcolsep}{4pt}  % 使列之间的空隙变窄

\resizebox{\textwidth}{!}{  % Scale the table to the text width
\begin{tabular}{ll|c|c|c|c|c|c|c}
\toprule
\multirow{2}{*}{\textit{SpeechLLMs}} & \multirow{2}{*} & \multicolumn{5}{c}{Closed-Ended} & \multicolumn{1}{|c|}{Open-Ended} & \multicolumn{1}{c}{Adjustment} \\ \cmidrule{3-4} \cmidrule{5-9}

& & \makecell{Standard \\ (P./I. Acc)$\uparrow$} & \makecell{Background\\(P./I. Acc)$\uparrow$} & \makecell{Accent\\(P./I. Acc)$\uparrow$} & \makecell{Paralinguistics\\(P./I. Acc)$\uparrow$} & \makecell{Disfluency\\(P./I. Acc)$\uparrow$} & \makecell{Standard\\(P./I. Acc)$\uparrow$} & \makecell{Standard\\(IAR$\uparrow$/ ECR$\downarrow$)} \\ \midrule

\multicolumn{9}{c}{\textit{English}} \\
\midrule

BLSP &  &  14.97 / 24.17 & 13.76 / 23.46 & 13.28 / 22.89 & 13.65 / 23.60 & 13.10 / 22.88 & 11.78 / 21.55 & 35.45 / 35.05 \\

GLM-4-Voice &  &  18.28 / 29.39 & 17.29 / 27.89 & 20.37 / 31.11 & 20.04 / 31.75 & 18.83 / 30.18 & 28.17 / 52.05 & 77.91 / 24.89 \\

Mini-Omni2 &  &  7.04 / 16.52 & 7.26 / 16.52 & 7.80 / 16.91 & 6.16 / 15.59 & 7.92 / 16.30 & 3.23 / 6.84 & 12.80 / 21.20 \\

Mini-Omni &  &  8.14 / 16.73 & 9.25 / 17.74 & 8.37 / 16.88 & 8.14 / 18.02 & 8.23 / 17.52 & 1.15 / 1.70 & 7.31 / 18.69 \\

Megrez &  &  19.49 / 31.04 & 17.51 / 28.96 & 18.31 / 29.18 & 19.02 / 30.53 & 19.28 / 28.67 & 37.27 / 64.35 & 55.06 / 31.98 \\

DIVA &  &  27.64 / 37.26 & 26.32 / 36.69 & 26.49 / 36.26 & 27.97 / 37.91 & 19.16 / 27.89 & 33.64 / 61.12 & 58.94 / 33.73 \\

Qwen2-Audio &  &  19.82 / 30.18 & 18.17 / 28.82 & 18.59 / 28.81 & 20.70 / 31.33 & 15.19 / 24.67 & 31.40 / 58.14 & 48.60 / 37.45 \\

\textbf{InSerter} & & \textbf{39.75} / \textbf{51.35} & \textbf{37.56} / \textbf{49.87} & \textbf{37.34} / \textbf{48.24} & \textbf{35.79} / \textbf{46.85} & \textbf{36.38} / \textbf{47.28} & \textbf{40.87} / \textbf{67.33} & \textbf{80.72} / \textbf{23.28} \\

\midrule
\multicolumn{9}{c}{\textit{Chinese}} \\
\midrule

GLM-4-Voice &  &  18.31 / 26.72 & 16.58 / 23.99 & 12.64 / 20.86 & 17.28 / 25.76 & 14.51 / 22.87 & 39.03 / 56.52 & 81.27 / 13.15 \\

% Mini-Omni2 &  &  12.44 / 21.82 & 10.82 / 20.22 & 10.98 / 19.51 & 11.75 / 20.38 & 10.59 / 20.22 & 0.14 / 0.17 & 0.42 / 0.64 \\

% Mini-Omni &  &  10.82 / 19.42 & 10.82 / 19.90 & 10.01 / 19.23 & 9.67 / 17.97 & 10.82 / 20.06 & 0.11 / 0.18 & 0.41 / 0.40 \\

Megrez &  &  18.31 / 27.84 & 17.16 / 26.32 & 17.26 / 26.01 & 18.43 / 27.92 & 16.58 / 25.76 & 31.69 / 39.27 & 63.34 / 17.53 \\

DIVA &  &  15.86 / 24.39 & 17.05 / 25.92 & 11.71 / 21.18 & 15.55 / 24.87 & 14.97 / 24.31 & 10.62 / 31.10 & 25.94 / 13.39 \\

Qwen2-Audio &  &  19.23 / 28.97 & 18.89 / 28.49 & 17.99 / 26.99 & 18.20 / 27.92 & 18.77 / 26.48 & 40.64 / 64.37 & 64.00 / 23.20 \\

\textbf{InSerter} &  &  \textbf{32.71} / \textbf{42.37} & \textbf{32.60} / \textbf{42.30} & \textbf{27.95} / \textbf{36.15} & \textbf{33.99} / \textbf{43.18} & \textbf{32.48} / \textbf{41.33} & \textbf{50.58} / \textbf{68.32} & \textbf{84.06} / \textbf{12.10} \\

\bottomrule
\end{tabular}
}  % End resizebox
\caption{Performance comparison of SpeechLLMs on SpeechInstructBench. P. and I. refer to prompt-level and instruction-level accuracy, respectively. For closed-ended questions, P.\&I. are calculated based on the average of both loose and strict accuracies. Adjustment tasks are evaluated using Instruction Adherence Rate (IAR) and Error Correction Rate (ECR). Notably, Mini-Omni2, Mini-Omni, and BLSP models are excluded from the Chinese benchmark comparisons due to their lack of Chinese language response capabilities.}
\label{tab:instruct-en}
% \vspace{2mm}
\end{table*}

\section{Experiments}

\subsection{Experimental Setup}
In this study, Qwen2-Audio-7B~\citep{Qwen2-Audio} is used as the base Language Model (LLM) for all experiments, where Whisper-Large-v3~\citep{radford2022whisper} encoder is used as the speech encoder to extract embeddings. The input speech segments are all resampled with a 16kHz sample rate and converted into 128-channel mel-spectrograms with a 25ms window and 10ms hop size, following a pooling layer with a stride of two.

For pre-training, we add 40\% speech-text interleaved data via InSerter, combined with other speech and text data. The model is trained with a global batch size of 1024 and the sequence length is fixed at 8192 tokens. Despite the increase in data volume, the model is trained for one epoch to ensure that the training amount on the original dataset is equivalent to that of the base model, facilitating a fair comparison. For supervised fine-tuning (SFT), we utilize the same dialogue dataset for all experiments to investigate the performance gains attributed only from InSerter pre-training. The fine-tuning process runs for one epoch. The Adam optimizer is used with a learning rate of $1e^{-5}$.

\subsection{Main Results}
\label{subsec:mainresults}

We evaluate InSerter on VoiceBench~\citep{chen2024voicebench} and SpeechInstructBench against several previous state-of-the-art SpeechLLMs, including Qwen2-Audio~\citep{Qwen2-Audio}, GLM-4-Voice~\citep{zeng2024glm4}, Mini-Omni~\citep{xie2024miniomnilanguagemodelshear}, Mini-Omni2~\citep{xie2024miniomni2opensourcegpt4ovision}, BLSP~\citep{wang2024blsp}, DIVA~\citep{held2024diva}, and Megrez-3B-Omni~\citep{megrez}.

\paragraph{VoiceBench results.}
\label{subsec:voicebench}

We adopt multiple evaluation tasks from VoiceBench~\citep{chen2024voicebench} to comprehensively evaluate our model's general speech processing capabilities. Across these diverse tasks shown in Table~\ref{tab:main-vb}, InSerter achieves competitive performance on multiple key metrics. In particular, our model demonstrates superior performance on AlpacaEval and CommonEval with GPT-scores of 4.23 and 3.63 in speech modality, respectively, and particularly excels on OpenBookQA with 77.14\%accuracy, substantially outperforming Qwen2-Audio (49.45\%) and other models. Additionally, InSerter maintains strong performance on benchmarks such as MMSU and AdvBench, showing its robust and balanced capabilities across various speech-related tasks. 

\paragraph{SpeechInstructBench results.}
\label{subsec:instructresult}

In SpeechInstructBench, evaluation metrics encompass prompt-level (P) and instruction-level (I) accuracy for both closed-ended and open-ended tasks. For closed-ended questions, P and I metrics are computed by averaging loose and strict accuracies, while adjustment task is assessed using Instruction Adherence Rate (IAR) and Error Correction Rate (ECR).

On the English benchmark (Table~\ref{tab:instruct-en}), InSerter achieves 39.75\% prompt-level accuracy and 51.35\% instruction-level accuracy on the standard closed-ended task, substantially surpassing the second-best performer DIVA (27.64\% and 37.26\%, respectively). InSerter maintains strong performance across all instruction-following tasks on SpeechInstructBench, especially on difficult scenarios (e.g., background noise), highlighting its robust capabilities. A similar pattern is observed in the Chinese benchmark (Table~\ref{tab:instruct-en}), where InSerter achieves state-of-the-art performance across all tasks. Notably, Mini-Omni2, Mini-Omni, and BLSP models are excluded from Chinese benchmark comparisons as they do not support Chinese language responses.

\begin{table}[!t]
\centering
\footnotesize
\vspace{-0.2cm}
\renewcommand{\arraystretch}{1.0}
\setlength{\tabcolsep}{2pt}  % Adjust column spacing
\begin{tabular}{lc|>{\centering\arraybackslash}p{2cm}}  % Use p{} to adjust only the ACC column width and center contents
\toprule
\multirow{2}{*}{\textit{Variations}} & \multirow{2}{*}{\textit{Training strategy}} & \multicolumn{1}{c}{Closed-Ended} \\ 
\cmidrule{3-3}
& & (P./I. Acc)$\uparrow$\\  % Ensure "test" is centered
\midrule

Baseline & W/o Inter. data & 23.40 / 31.75 \\

BLSP & Continuation Writing& 28.20 / 39.27 \\

DIVA & Distillation & 26.81 / 37.58 \\

\textbf{InSerter} & Sentence-Level & 31.08 / 42.98 \\

& Word-Level & \textbf{36.56} / \textbf{47.38} \\

\midrule

\textbf{InSerter}  & + Distillation & 35.56 / 46.38 \\

& + CW. \& Distillation & 38.78 / 50.96 \\

& + Continuation Writing & \textbf{39.75} / \textbf{51.35} \\

\bottomrule
\end{tabular}
\caption{We compare different alignment strategies, where "BLSP" refers to using the same alignment method as BLSP~\citep{wang2024blsp}, "DIVA" denotes the DIVA~\citep{held2024diva} alignment method, and "CW." represents continuation writing data. For our method, the pre-training experiments are conducted at both the word-level and sentence-level interleaved data. Speech instruction-following performance is evaluated using the closed-ended SpeechInstructBench dataset (English standard version).}
\label{tab:method-compare}
\end{table}

\begin{figure*}[htbp]
    \centering
    % First subfigure
    \begin{minipage}{0.24\textwidth}
        \centering
        \includegraphics[width=\textwidth]{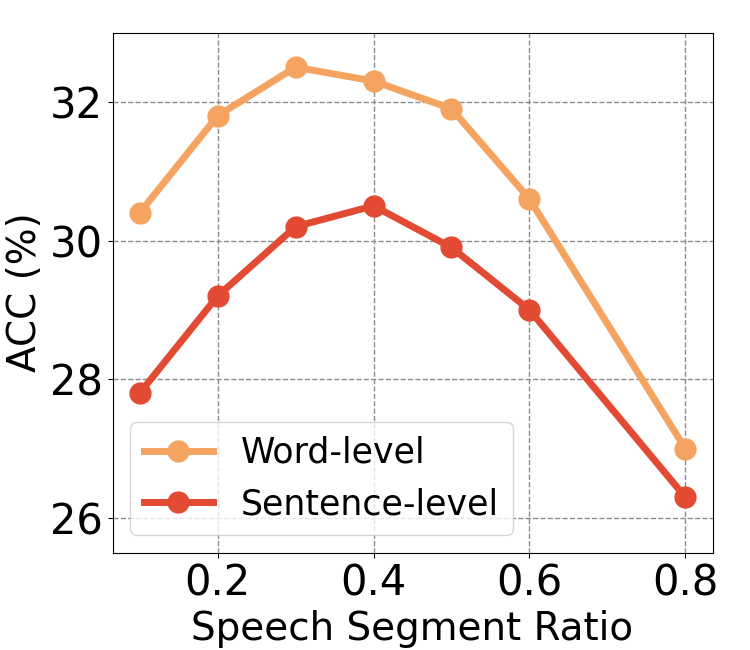}  % Replace with your file
        \subcaption{Speech segments ratio} \label{fig:subfig1}
    \end{minipage} \hfill % Adjust horizontal space
    % Second subfigure
    \begin{minipage}{0.24\textwidth}
        \centering
        \includegraphics[width=\textwidth]{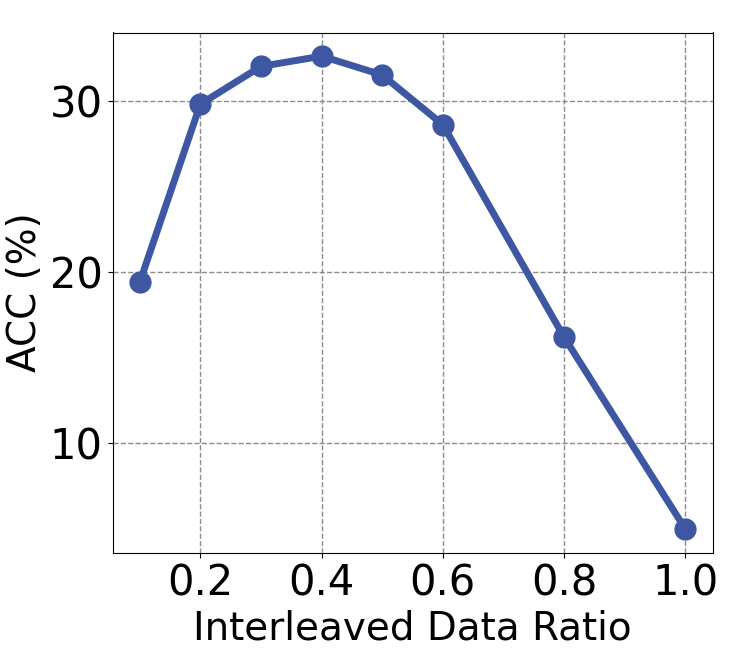}  % Replace with your file
        \subcaption{Interleaved data ratio} \label{fig:subfig2}
    \end{minipage} \hfill % Adjust horizontal space
    % Third subfigure
    \begin{minipage}{0.24\textwidth}
        \centering
        \includegraphics[width=\textwidth]{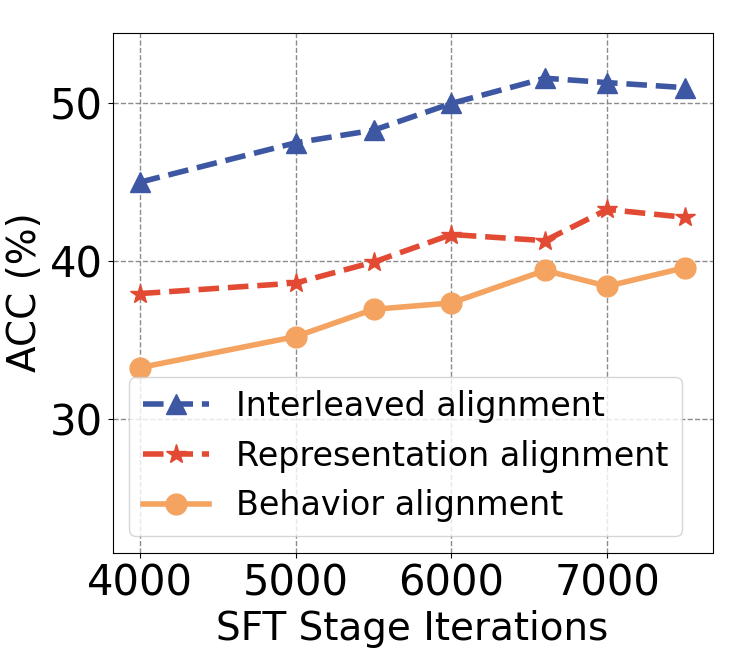}  % Replace with your file
        \subcaption{Iterations trend} \label{fig:subfig3}
    \end{minipage} \hfill % Adjust horizontal space
    % Fourth subfigure
    \begin{minipage}{0.24\textwidth}
        \centering
        \includegraphics[width=\textwidth]{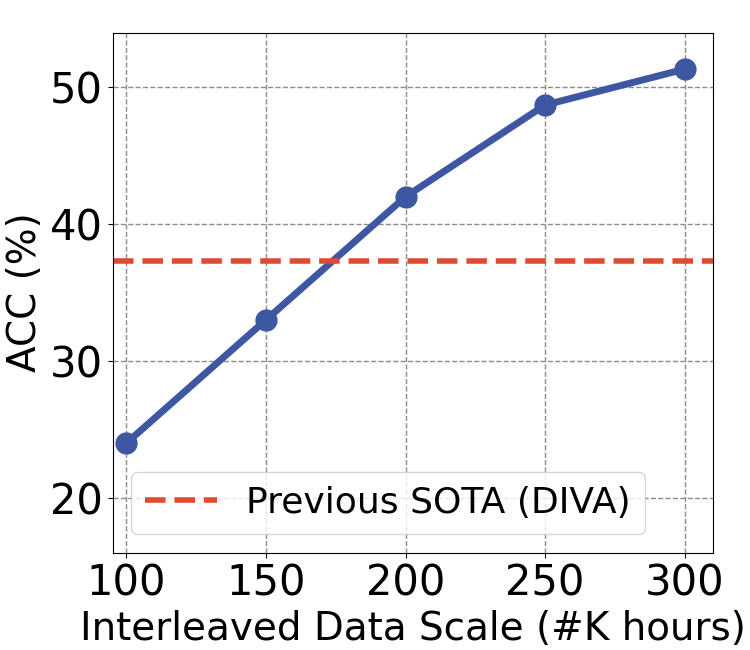}  % Replace with your file
        \subcaption{Interleaved data scale} \label{fig:subfig4}
    \end{minipage}

    % Main caption for the whole figure
    \caption{(a) Performance variation with different speech segment ratios. (b) Impact of interleaved data proportion on performance. (c) Different alignment strategies performance analysis with iterations during the SFT stage. (d) Performance benefits from expanding data scale. All accuracy scores refer to the closed-ended SpeechInstructBench (English standard-version) instruction-level accuracy.}
    \label{fig:full_figure}
    \vspace{-4mm}
\end{figure*}

\subsection{Ablation Study}
\subsubsection{Alignment Strategy Comparison}
\label{subsec:compare}

To ensure a fair comparison of different post-training alignment algorithms (behavior alignment~\citep{wang2024blsp, held2024diva} and representation alignment~\citep{held2024diva}), we develop a baseline pre-training checkpoint derived from Qwen2-Audio. This baseline is trained without interleaved data, while maintaining the total dataset volume and pre-training model configuration consistent with InSerter. During the supervised fine-tuning (SFT) phase, in addition to training on the same dialogue data as InSerter, we incorporate different post-training alignment methods' training tasks and data. The mixing ratio is carefully adjusted, with experiments conducted at intervals of 10\% ranging from 10\% to 90\%, to ensure that these algorithms achieve optimal performance. Unless otherwise specified, we employ the same settings for all baselines and InSerter, including training hyperparameters. For post-training datasets, we construct the same amount of data for BLSP~\citep{wang2024blsp} and DIVA~\citep{held2024diva} experiments to match their original implementations: 8.8 million (speech, text continuation) pairs from GigaSpeech\citep{GigaSpeech2021} and Wenetspeech\citep{zhang2022wenetspeech} datasets for BLSP, and 3.5 thousand hours of CommonVoice 17~\citep{ardila-etal-2020-common} data for DIVA. 

Table~\ref{tab:method-compare} demonstrates InSerter's superior performance when using word-level interleaved data. Compared to existing alignment methods and the baseline, it achieves significant improvements of 36.56\% in prompt-level accuracy and 47.38\% in instruction-level accuracy. The word-level InSerter approach proves more effective compared with sentence-level as it aligns better with the text continuation objective, providing finer granularity for speech-text alignment. Based on our ablation studies (Fig.~\ref{fig:full_figure}(a)), we determined the optimal speech segment ratios to be 30\% for word-level and 40\% for sentence-level interleaved data.

Moreover, model performance gains are further enhanced when combining interleaved pre-training strategy with existing post-training alignment methods. Specifically, integrating word-level interleaved pre-training method with continuation writing post-training strategy yields additional improvements of 39.56\% (prompt-level accuracy) and 51.35\% (instruction-level accuracy). We use this version as the final competitive model for the main results section (Sec.~\ref{subsec:mainresults}).

\subsubsection{Hyperparameters and Data Scaling Analysis}
\label{subsec:ablation}

To rigorously evaluate the effectiveness of our InSerter approach, we study the effects of data scaling and InSerter related hyperparameters. First, as shown in Fig.~\ref{fig:full_figure}(a), we conduct experiments on the pre-training stage to determine the optimal speech segment ratio for interleaved data. The results reveal that model performance peaks when speech segment ratio falls between 20\% and 40\%, with the highest accuracy achieved around 30\% for word-level and 40\% for sentence-level interleaved data. As presented in Fig.~\ref{fig:full_figure}(b), varying the interleaved data ratio among pre-training data greatly influences performance, with an optimal ratio nearing 0.4 for achieving the highest accuracy. Fig.~\ref{fig:full_figure}(c) presents the impact of iterations during the supervised fine-tuning (SFT) stage, indicating that interleaved alignment outperforms representation and behavior alignments consistently over increasing iterations, peaking between 6000 and 7000 iterations. So we set SFT with 7000 iterations for all experiments. Finally, Fig.~\ref{fig:full_figure}(d) illustrates the relationship between interleaved data scale and instruction-following performance, confirming that larger interleaved data scales lead to substantial performance improvements, ultimately surpassing the previous state-of-the-art accuracy (as indicated by the red dashed line for DIVA~\citep{held2024diva}) when data scare reaches around 300K hours. This proves that our proposed method can scale effectively and yield more performance benefits by continuously expanding the interleaved data.

\section{Conclusion}

In this paper, we introduced a novel and easily scalable pre-training framework termed InSerter, alongside a comprehensive speech instruction-following benchmark. Our proposed InSerter demonstrates state-of-the-art performance across various instruction-following tasks while maintaining competitive results on diverse speech-language tasks. Future work would explore more efficient training techniques and investigate the potential of extending our approach to more diverse languages and speech scenarios.

\section{Limitation}

Despite the promising results obtained in our method, there are still several areas for improvement. Firstly, the training process consists of only English and Chinese related datasets. The generalization to other languages is yet to be verified. Secondly, for evaluation, some tasks of SpeechInstructBench (open-ended instruction, adjustment instruction) heavily rely on GPT-4o API for scoring. However, the availability and accessibility of the GPT-4o API are external factors beyond our control. Addressing these limitations in future work would provide a more comprehensive validation of our approach.

\section{Ethical Consideration}

We affirm that we contribute to society, avoid harm, and are honest and trustworthy. We respect previous work and appropriately cite the methods and datasets we are using. All the data we use is subject to strict verification to ensure it contains no harmful or sensitive content.

\bibliography{custom}
% \newpage
\appendix

\section{Pre-Train Datasets}
\label{subsec:pertraindata}

Table~\ref{table_chat_statistic} shows the statistics of the datasets used for the pre-training of InSerter. The data comprises three types: interleaved speech-text data, unsupervised text data, and multi-task speech data. These large-scale pre-training datasets are primarily collected from sources such as podcasts, MOOCs, articles, webpages, Wikipedia, and books. The interleaved speech-text data includes datasets like SpokenWOZ, while the multi-task speech data includes collected multi-task speech datasets following the Qwen-Audio series~\citep{Qwen-Audio,Qwen2-Audio}. The combination of these rich and diverse data sources facilitates effective pre-training for InSerter.

\section{SpeechInstructBenchmark Design}
\label{subsec:appendspeechinstructbenchdesign}
\subsection{Benchmark Data Statistics}
Table~\ref{stat_bench} presents the data statistics of SpeechInstructBench, categorized into closed-ended, open-ended, and adjustment types. The datasets are sourced from IFEval~\citep{zhou2023instructionfollowingevaluationlargelanguage}, FollowBench~\citep{jiang-etal-2024-followbench}, InFoBench~\citep{qin2024infobench}, Alpaca~\citep{alpaca}, and InfinityInstruct~\citep{InfinityInstruct2024}, spanning both English (En) and Chinese (Cn) languages. The Chinese data is translated from the English version and has undergone careful manual review. 

% paper-(type, tasks, example, num)
\subsection{Prompts for Creating Data and Evaluation}

In this part, we partially demonstrate the process of adjusting the prompt aimed at assessing the instruction-following performance of models.

\begin{figure*}[!t]
    \centering
    \includegraphics[width=1\linewidth]{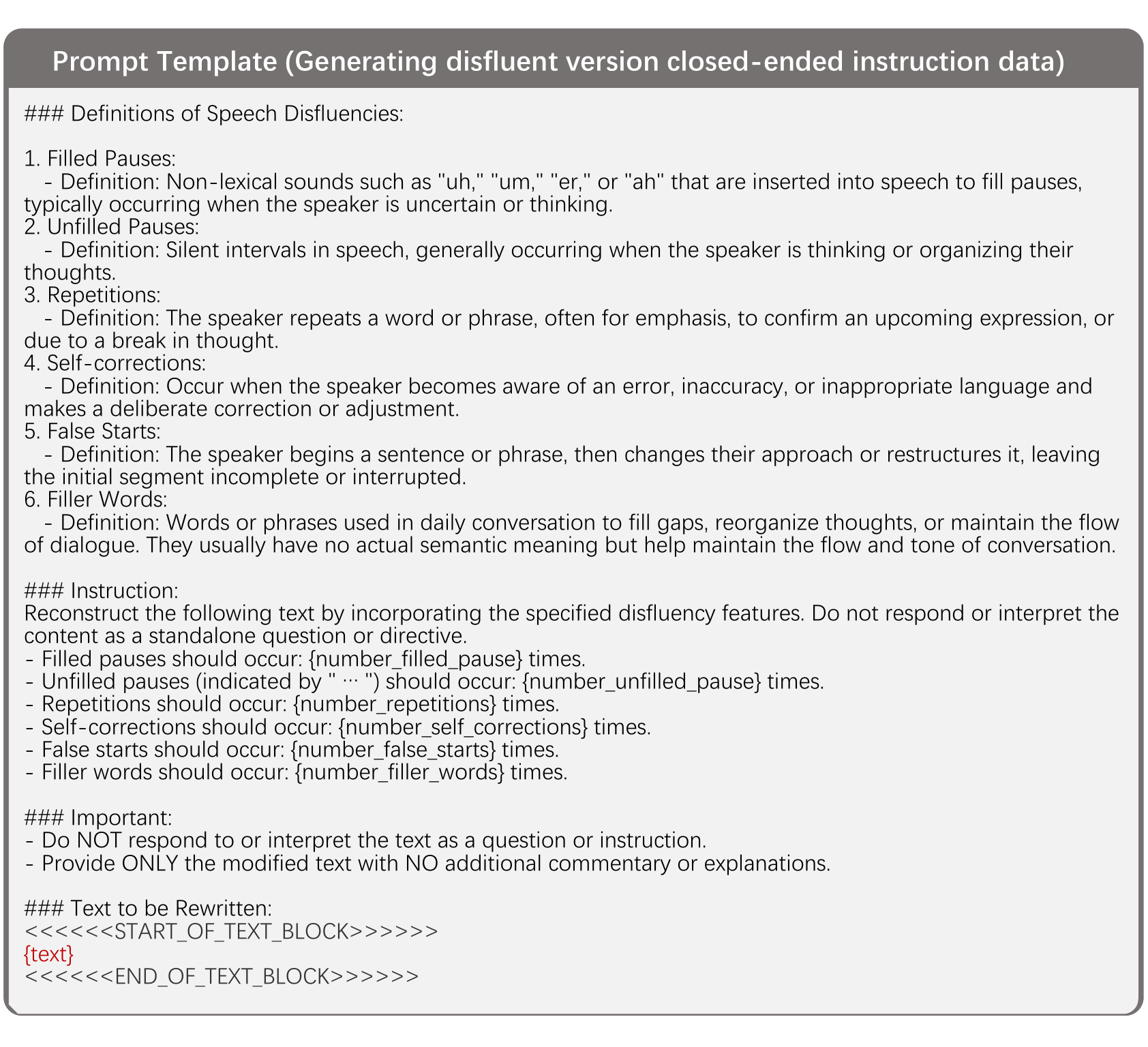}
    \caption{GPT prompt for generating disfluent version closed-ended instruction data.}
    \label{fig:disfluency}
\end{figure*}

When constructing disfluency instruction data (as shown in Fig.~\ref{fig:disfluency}), it is crucial to include specific prompts, such as "Do NOT respond to or interpret the text as a question or instruction. Provide ONLY the modified text with NO additional commentary or explanations." Without these explicit instructions, GPT can sometimes generate responses that address the original input's instructions instead of simply creating the disfluent text. Moreover, the understanding of certain definitions, like the construction of adjustment instruction data, benefits from the inclusion of descriptive definitions and concrete examples within the prompt. As adjustment instruction data represents a novel task, providing detailed guidance within the prompt enables GPT to produce outputs that align more closely with the desired criteria.

\begin{figure*}[!t]
    \centering
    \includegraphics[width=1\linewidth]{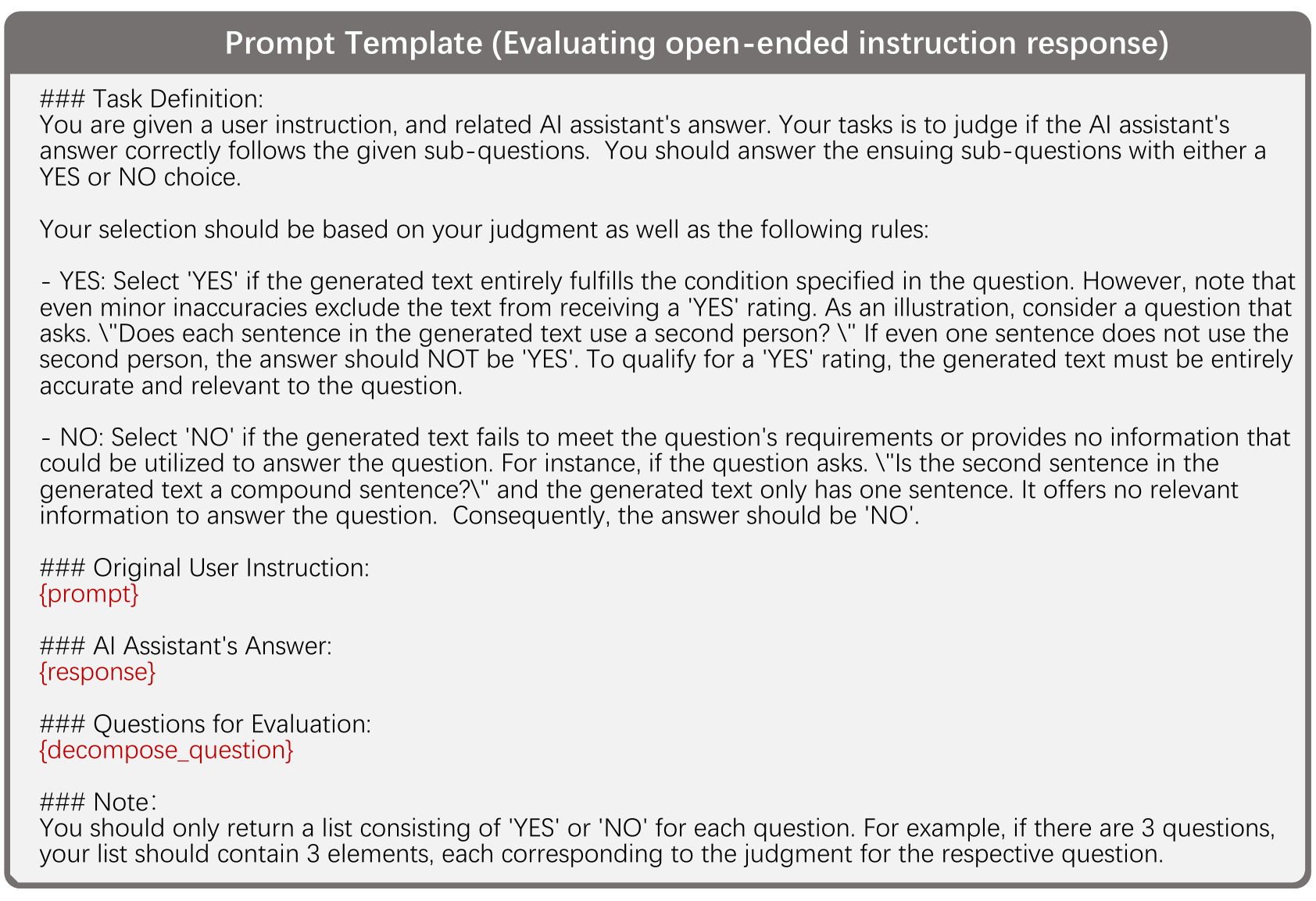}
    \caption{GPT prompt for evaluating open-ended instruction task.}
    \label{fig:openeval}
\end{figure*}

\begin{figure*}[!t]
    \centering
    \includegraphics[width=1\linewidth]{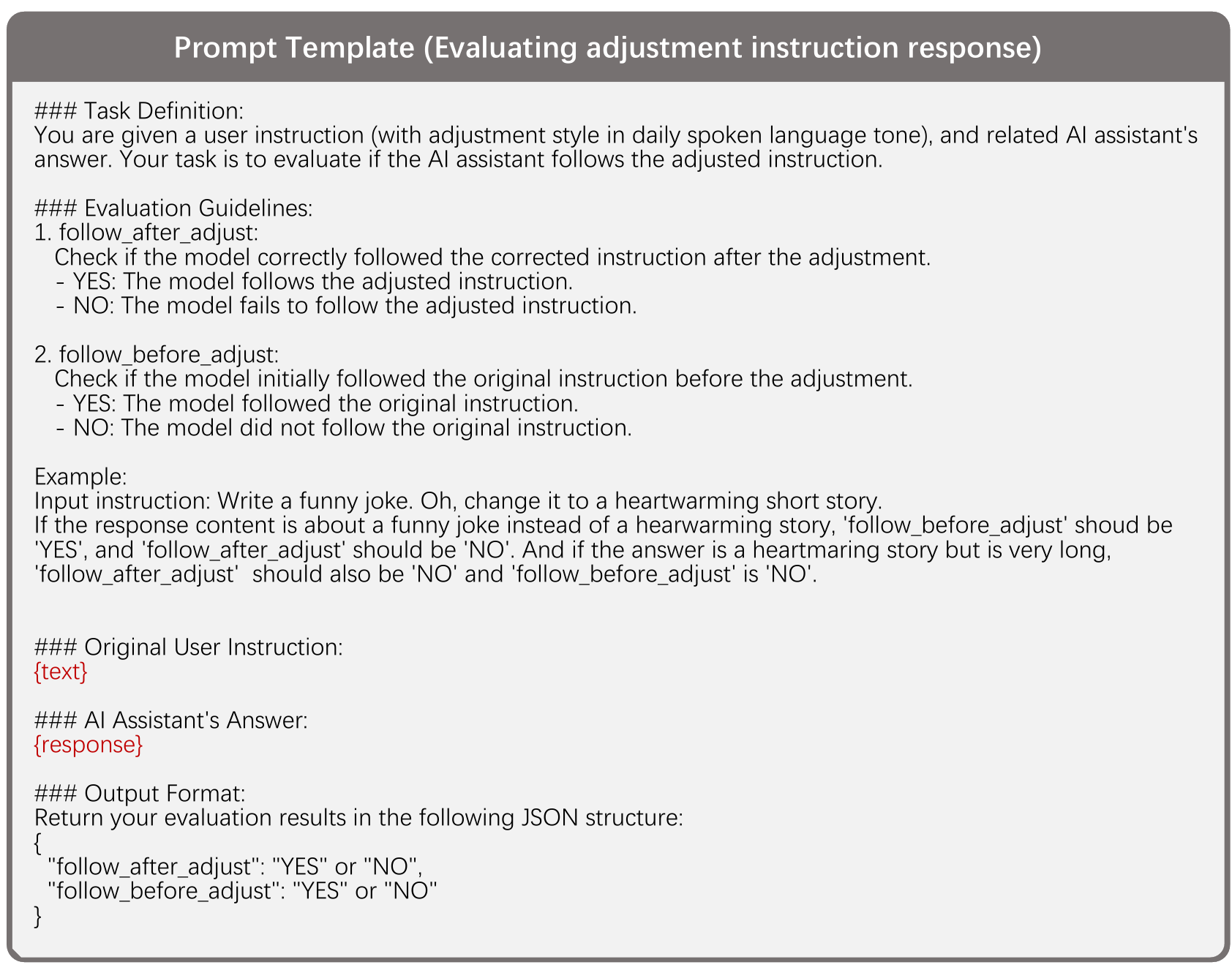}
    \caption{GPT prompt for evaluating adjustment instruction task.}
    \label{fig:adjusteval}
\end{figure*}

For GPT prompts for evaluation (Fig.~\ref{fig:openeval}, Fig.~\ref{fig:adjusteval}), we observed that excessively long prompt content can occasionally lead to the model not fully adhering to the task requirements. As a result, when designing evaluation prompts, we focused on streamlining the expression of task definitions and requirements. This was achieved by using concise language and structuring the prompt with clear subdivisions (e.g., using "\#\#\#\ " to separate each section), which helps in maintaining clarity and focus. Furthermore, we emphasized providing explicit output format requirements, such as specifying that the answer should be "YES" or "NO". This structured approach not only reduces the cognitive load on the model but also enhances its ability to comply with the task criteria, resulting in more reliable and accurate evaluations. By refining the prompt design in this manner, we aim to improve the overall performance of GPT in tasks involving instruction comprehension and response evaluation.

\subsection{Human Assess of SpeechInstructBench}

\begin{figure*}[!t]
    \centering
    \includegraphics[width=1\linewidth]{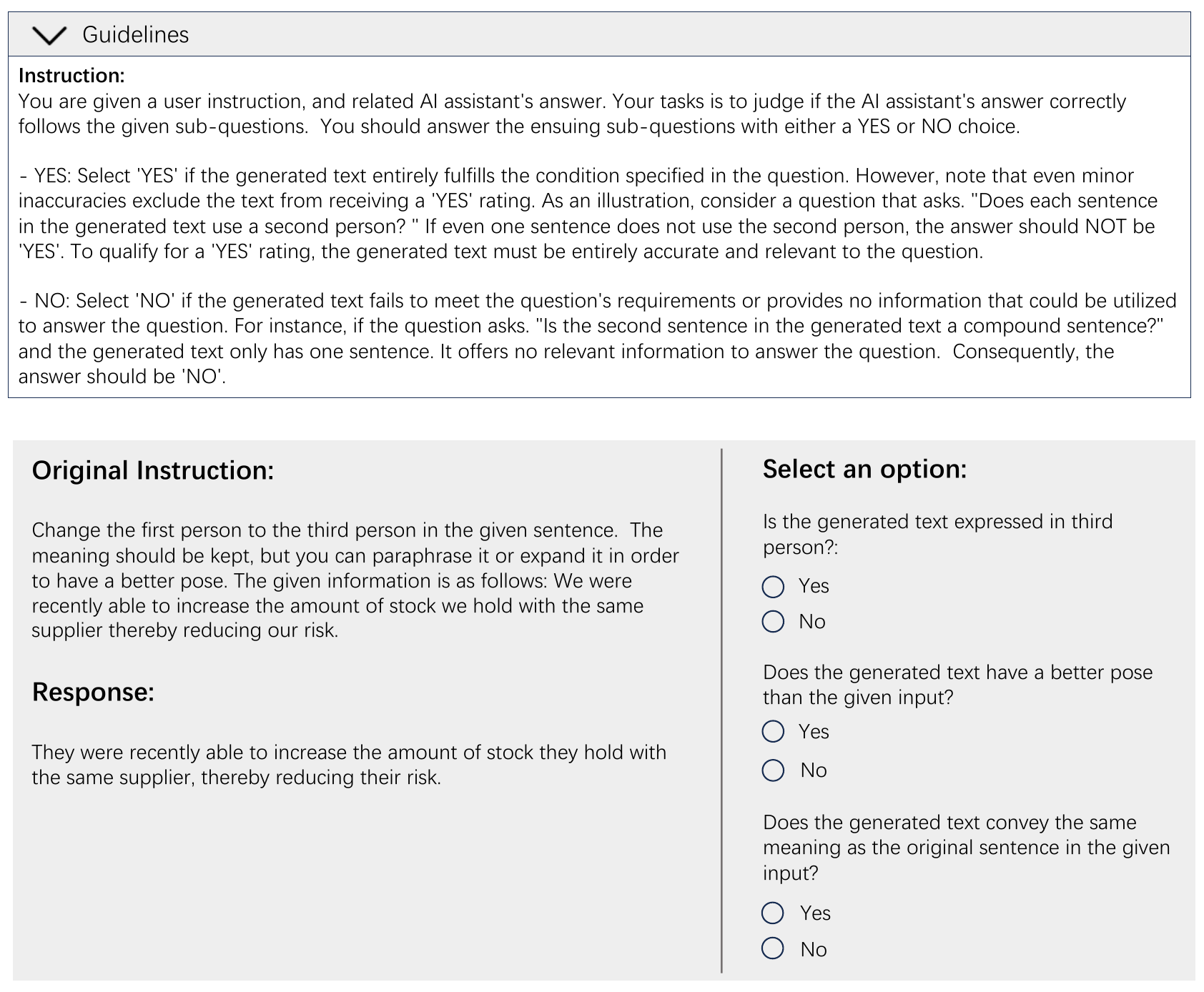}
    \caption{Screenshot of human evaluation for the open-ended instruction task.}
    \label{fig:useropen}
\end{figure*}

\begin{figure*}[!t]
    \centering
    \includegraphics[width=1\linewidth]{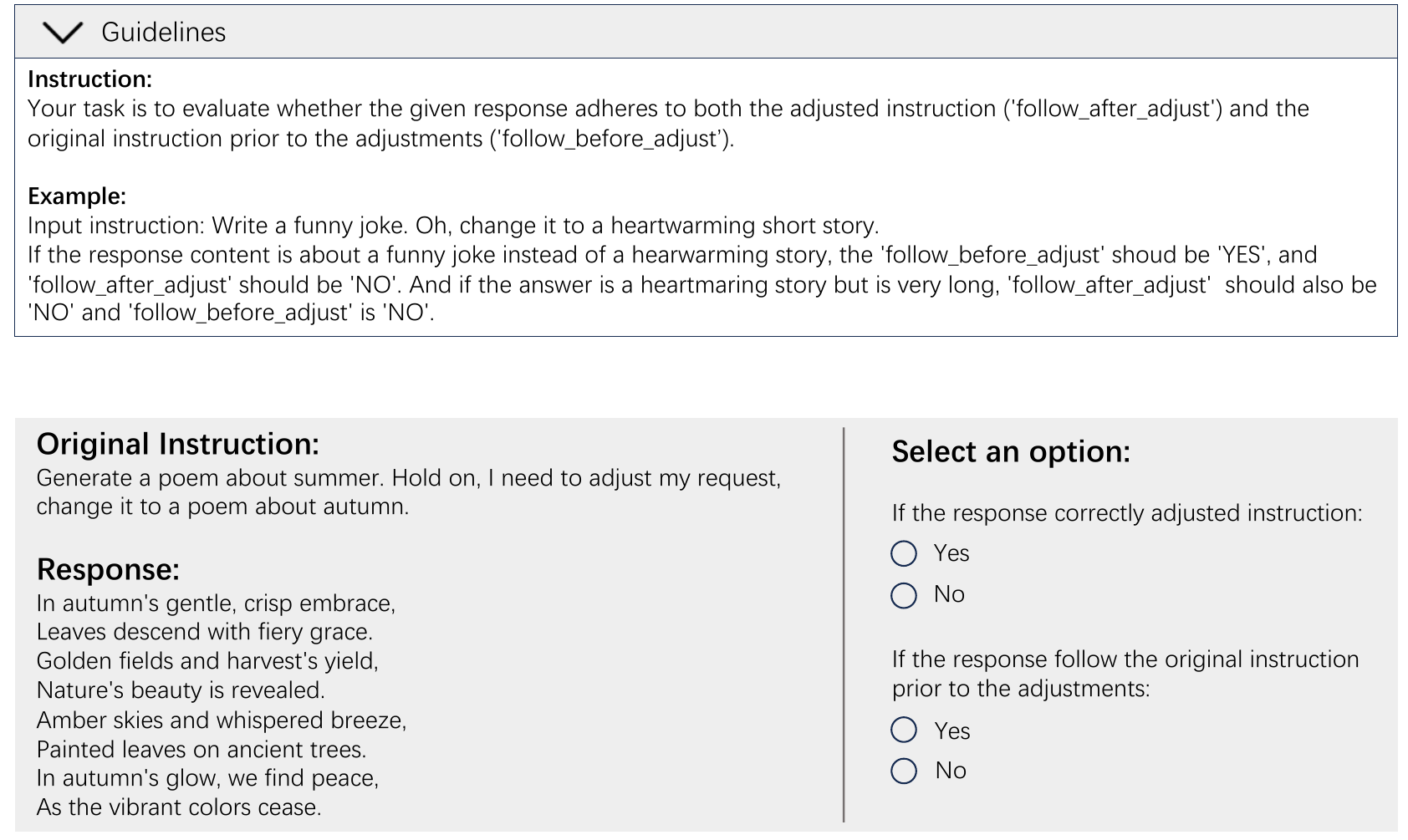}
    \caption{Screenshot of human evaluation for the adjustment instruction task.}
    \label{fig:useradjust}
\end{figure*}

To evaluate the effectiveness of the GPT-4o scoring mechanism within SpeechInstructBench, we conducted a human evaluation process. A total of 20 evaluators participated in this assessment, where they randomly reviewed 100 items from the dataset. This selection includes 50 open and closed questions (25 in English and 25 in Chinese) and 50 adjustment items (25 in English and 25 in Chinese). As shown in Fig.~\ref{fig:useropen} and Fig.~\ref{fig:useradjust}, the screenshot captures the process of human evaluation. During evaluation, these evaluators assessed the factual accuracy of each response. They were required to make their assessments by selecting either "YES" or "NO" to indicate whether the response was factually correct.

\section{Case Study}

\begin{figure*}[!t]
    \centering
    \includegraphics[width=1\linewidth]{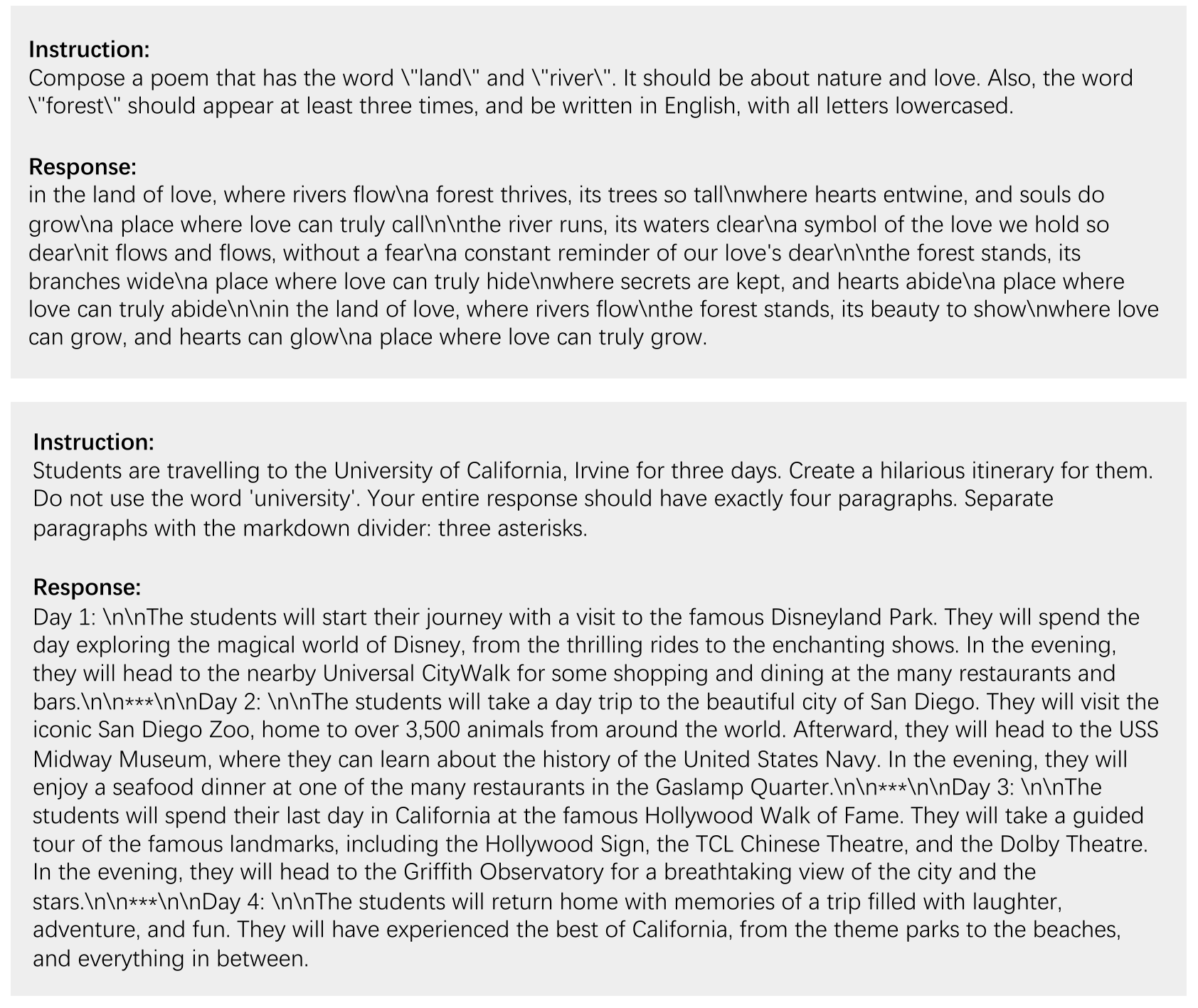}
    \caption{SpeechInstructBench Task: Closed-ended instruction following (English version)}
    \label{fig:democloseen}
\end{figure*}

\begin{figure*}[!t]
    \centering
    \includegraphics[width=1\linewidth]{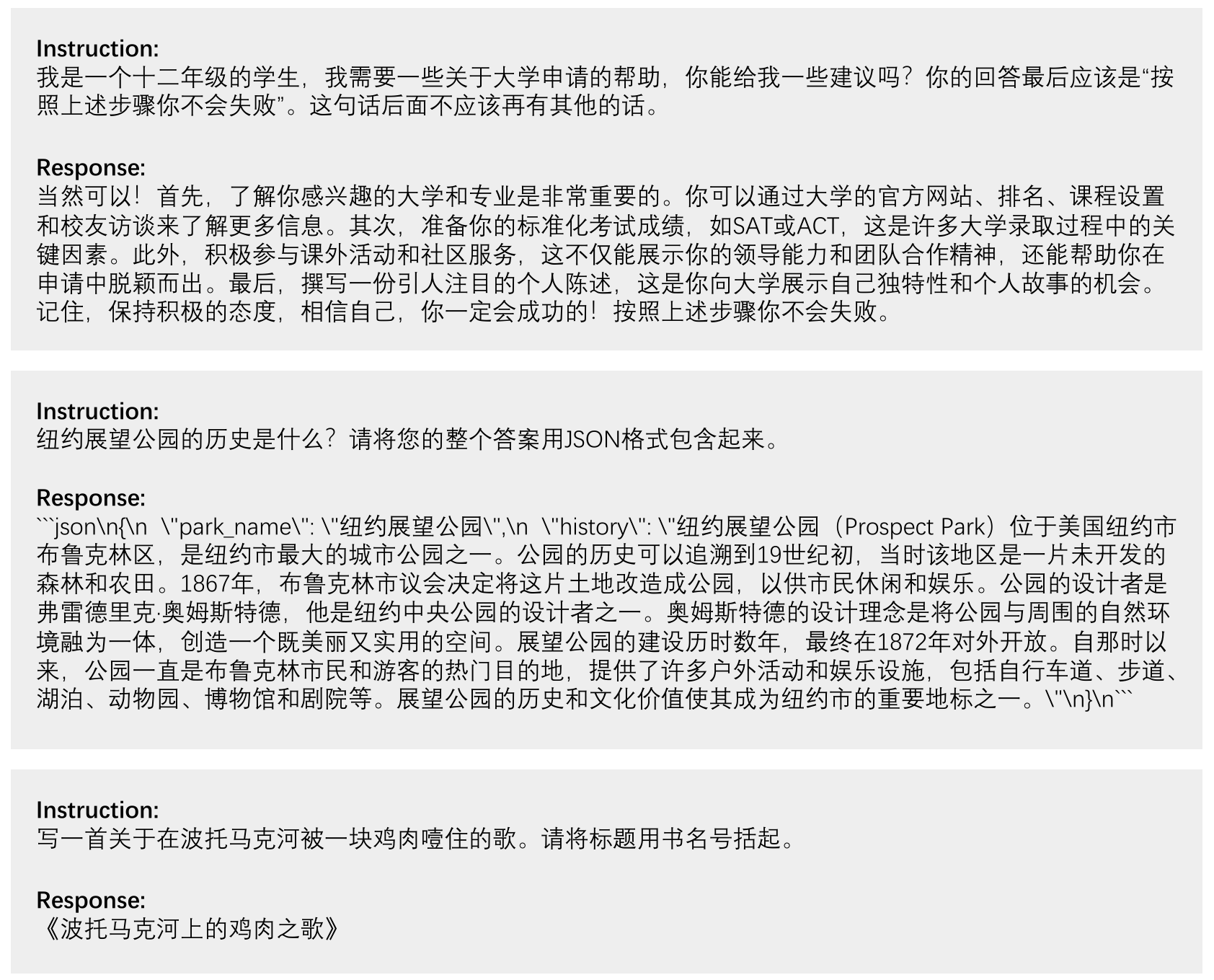}
    \caption{SpeechInstructBench Task: Closed-ended instruction following (Chinese version)}
    \label{fig:democlosecn}
\end{figure*}

\begin{figure*}[!t]
    \centering
    \includegraphics[width=1\linewidth]{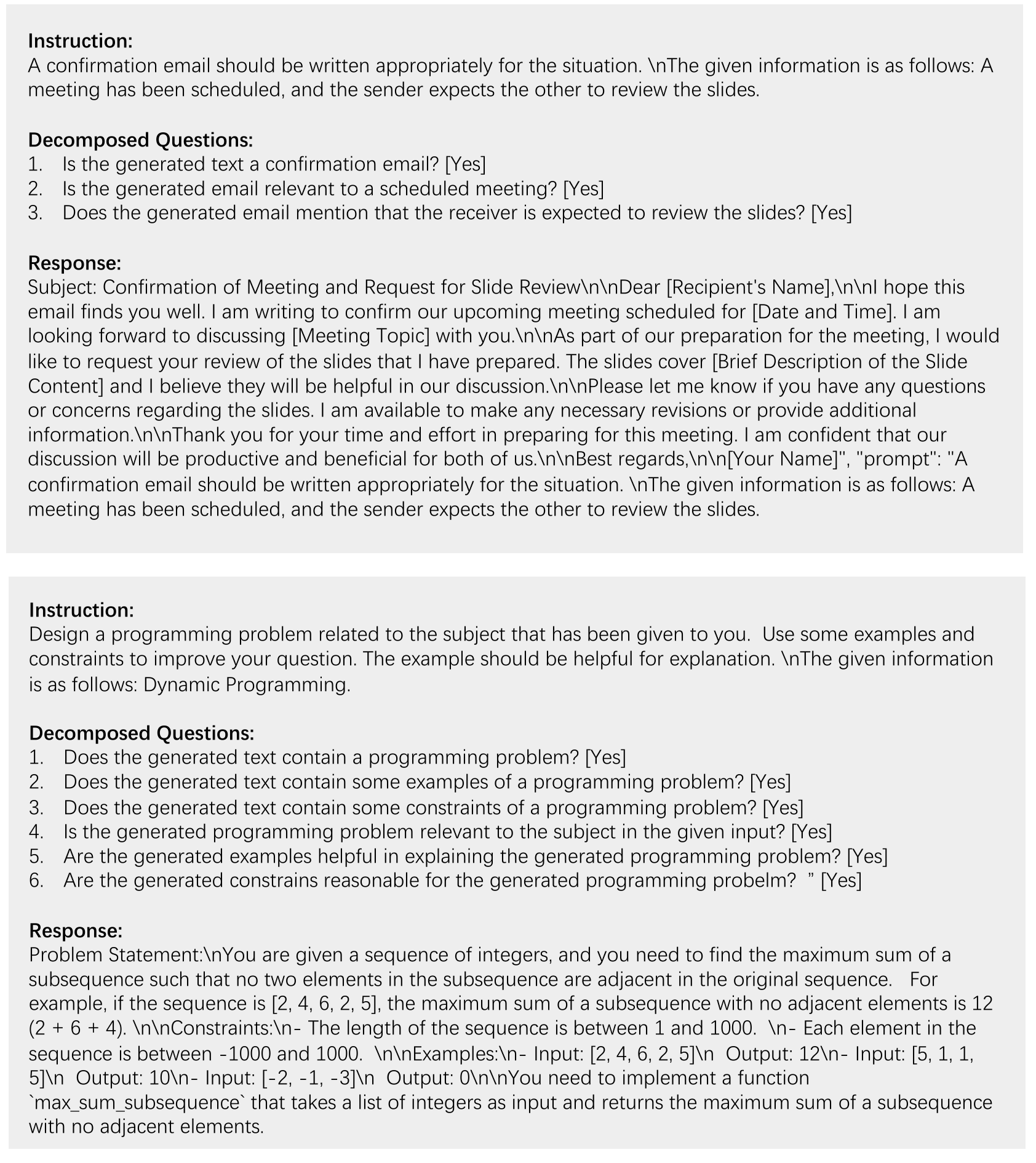}
    \caption{SpeechInstructBench Task: Open-ended instruction following - English version}
    \label{fig:demoopenen}
\end{figure*}

\begin{figure*}[!t]
    \centering
    \includegraphics[width=1\linewidth]{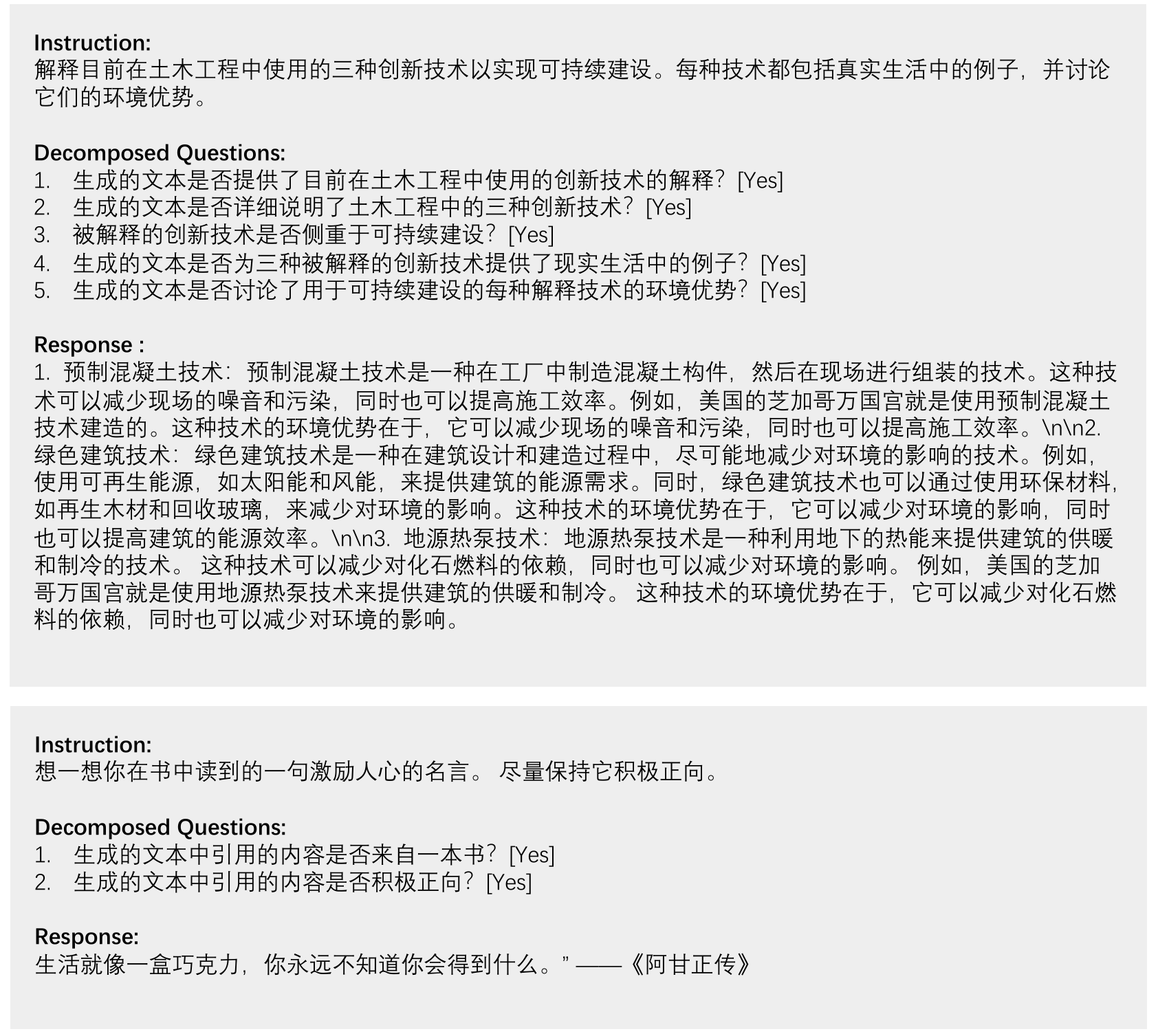}
    \caption{SpeechInstructBench Task: Open-ended instruction following (Chinese version)}
    \label{fig:demoopencn}
\end{figure*}

\begin{figure*}[!t]
    \centering
    \includegraphics[width=1\linewidth]{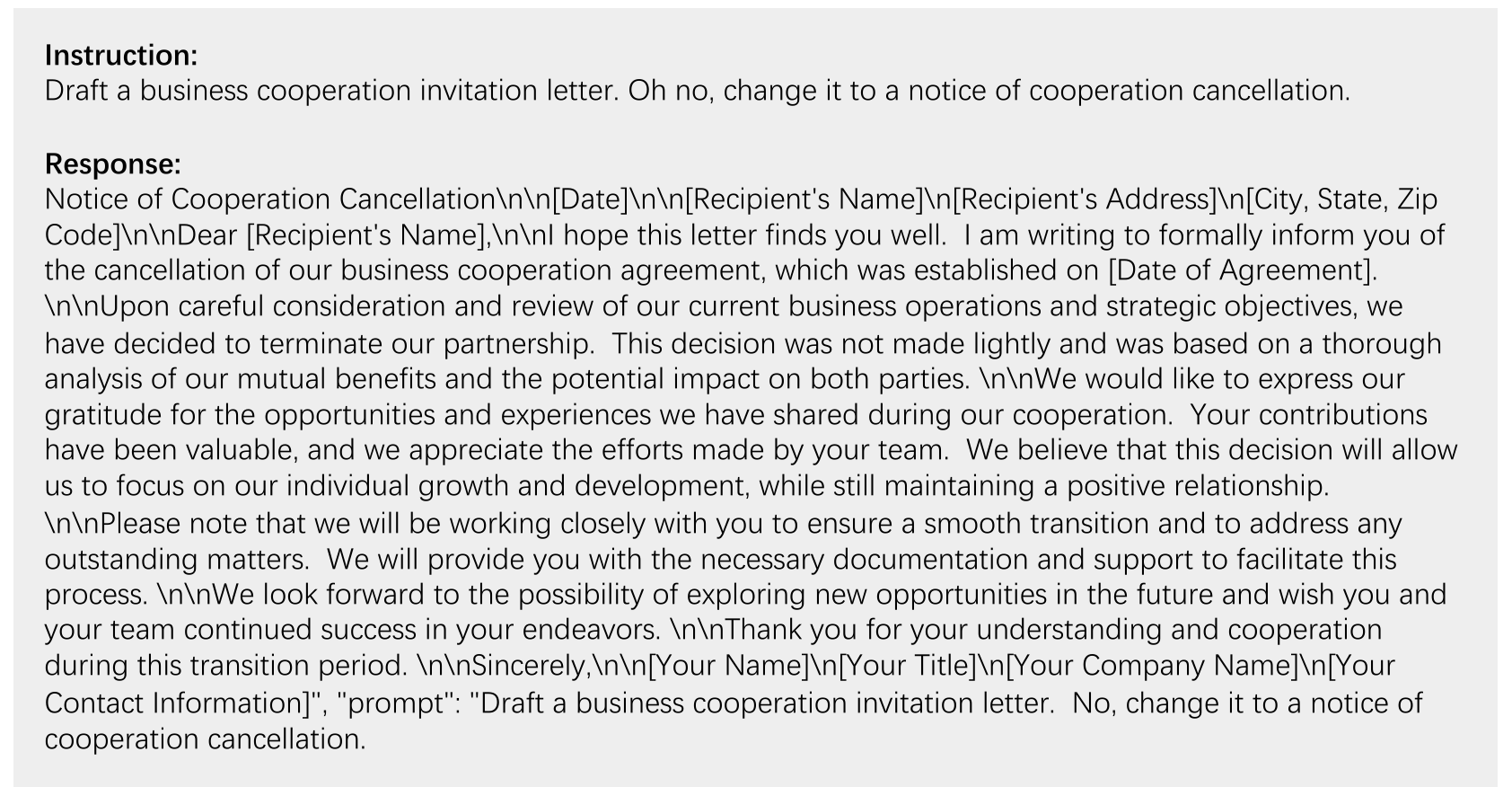}
    \caption{SpeechInstructBench Task: Adjustment instruction following (English version)}
    \label{fig:demoadjusten}
\end{figure*}

\begin{figure*}[!t]
    \centering
    \includegraphics[width=1\linewidth]{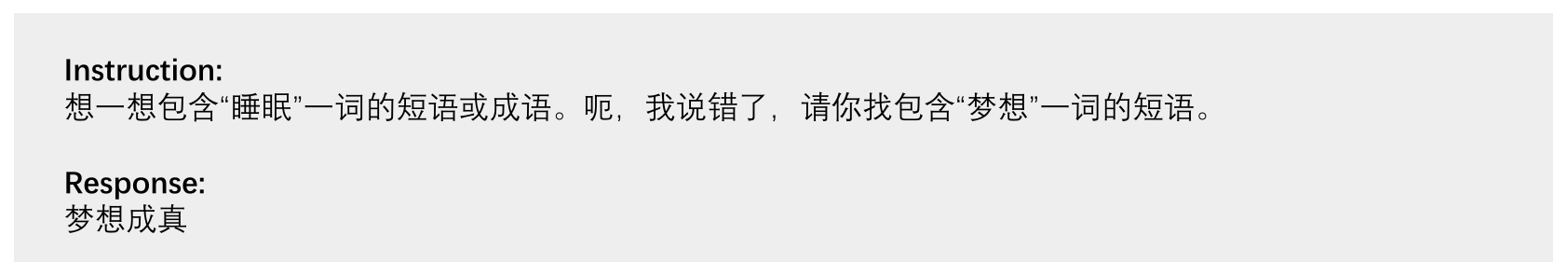}
    \caption{SpeechInstructBench Task: Adjustment instruction following (Chinese version)}
    \label{fig:demoadjustcn}
\end{figure*}

\begin{figure*}[!t]
    \centering
    \includegraphics[width=1\linewidth]{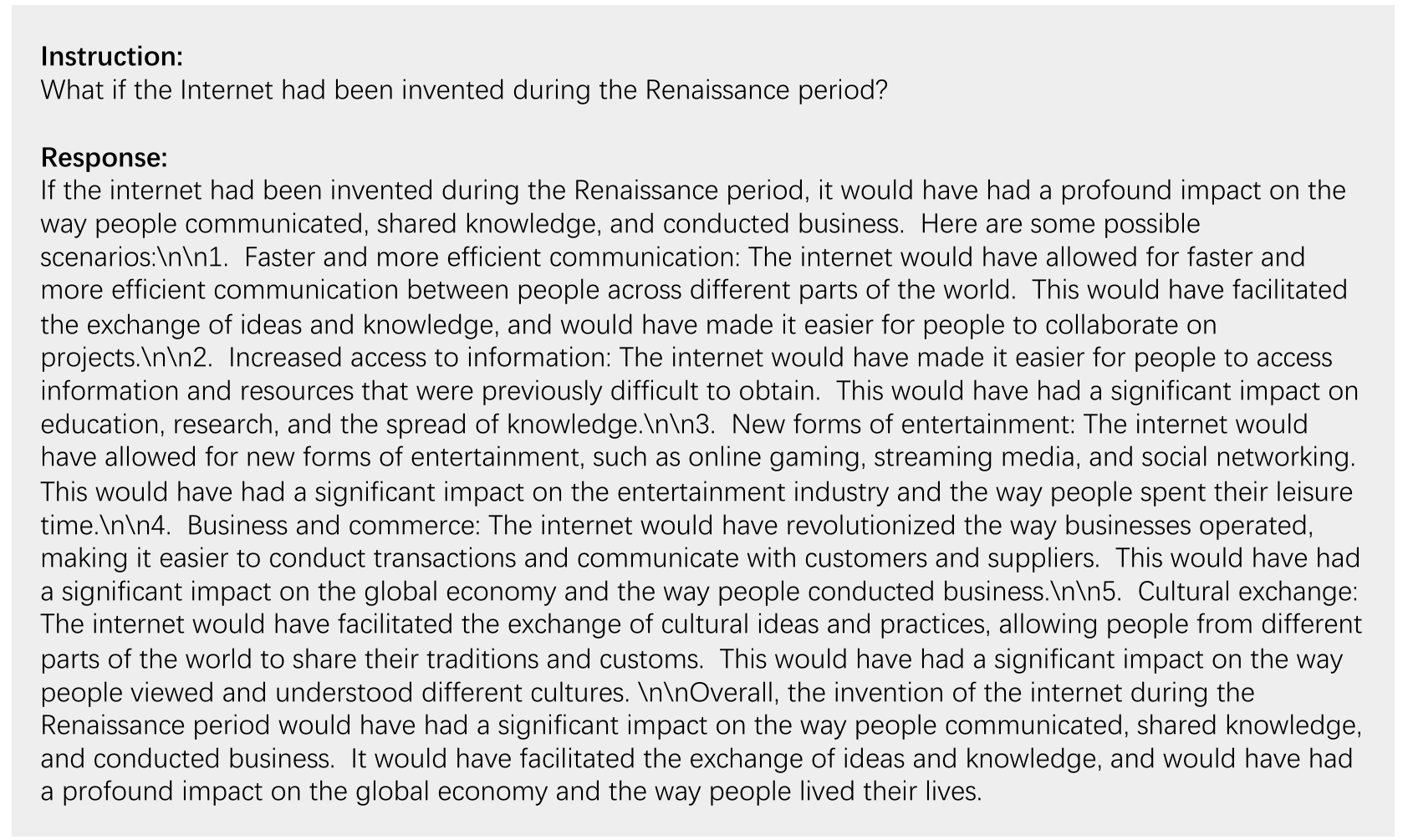}
    \caption{VoiceBench Task: AlpacaEval}
    \label{fig:alpacaeval}
\end{figure*}

\begin{figure*}[!t]
    \centering
    \includegraphics[width=1\linewidth]{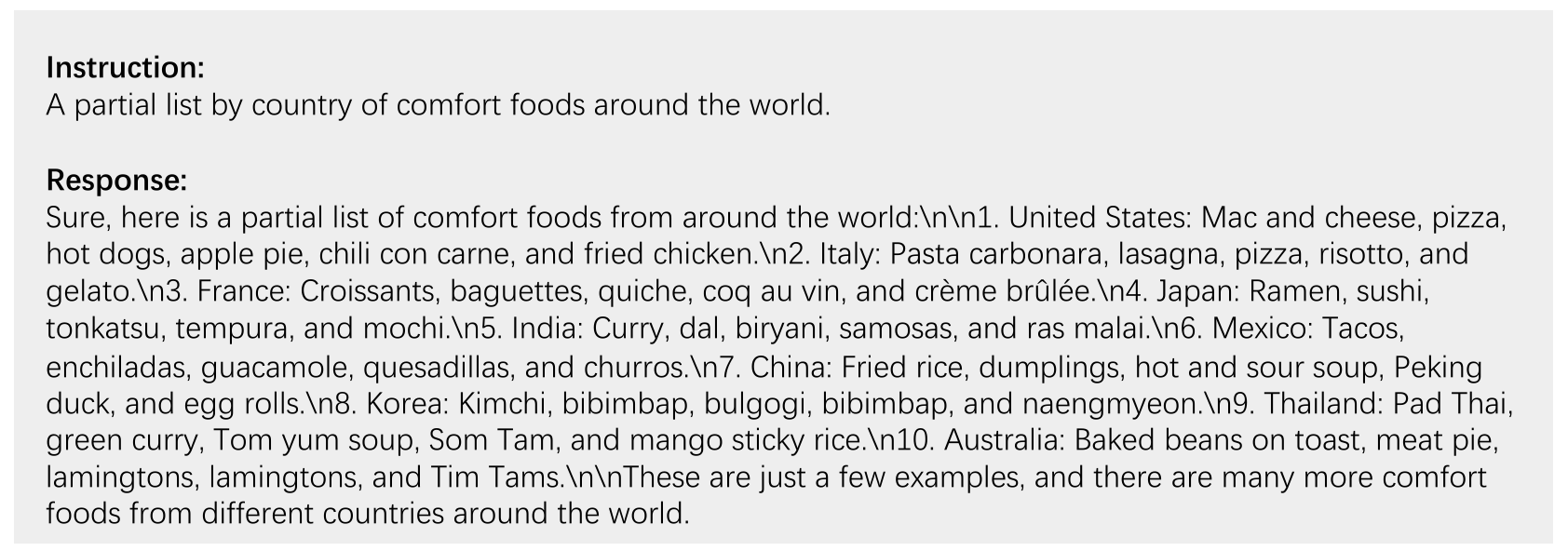}
    \caption{VoiceBench Task: CommonEval}
    \label{fig:commoneval}
\end{figure*}

\begin{figure*}[!t]
    \centering
    \includegraphics[width=1\linewidth]{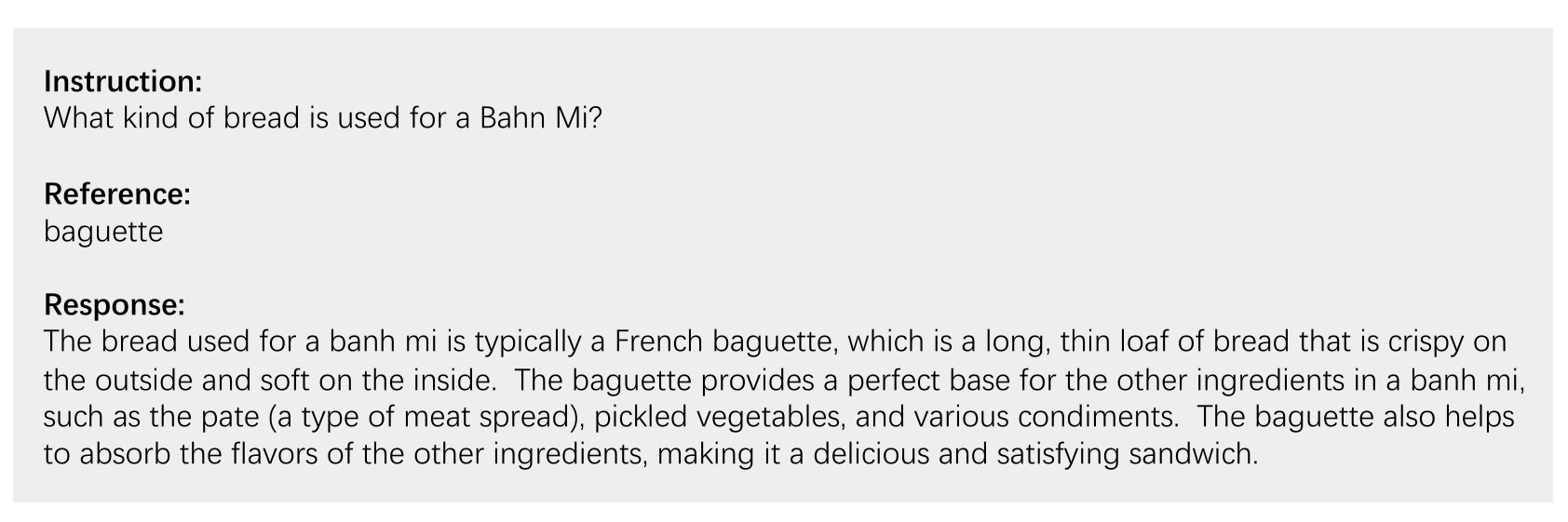}
    \caption{VoiceBench Task: SD-QA}
    \label{fig:sdqa}
\end{figure*}

\begin{figure*}[!t]
    \centering
    \includegraphics[width=1\linewidth]{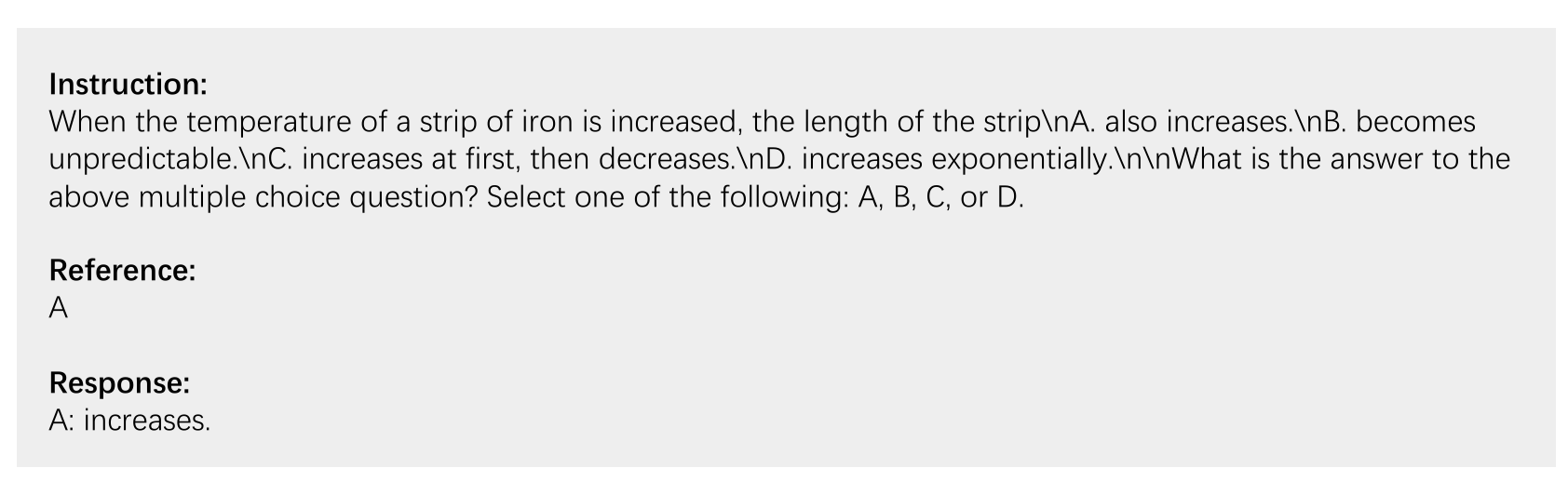}
    \caption{VoiceBench Task: MMSU}
    \label{fig:mmsu}
\end{figure*}

\begin{figure*}[!t]
    \centering
    \includegraphics[width=1\linewidth]{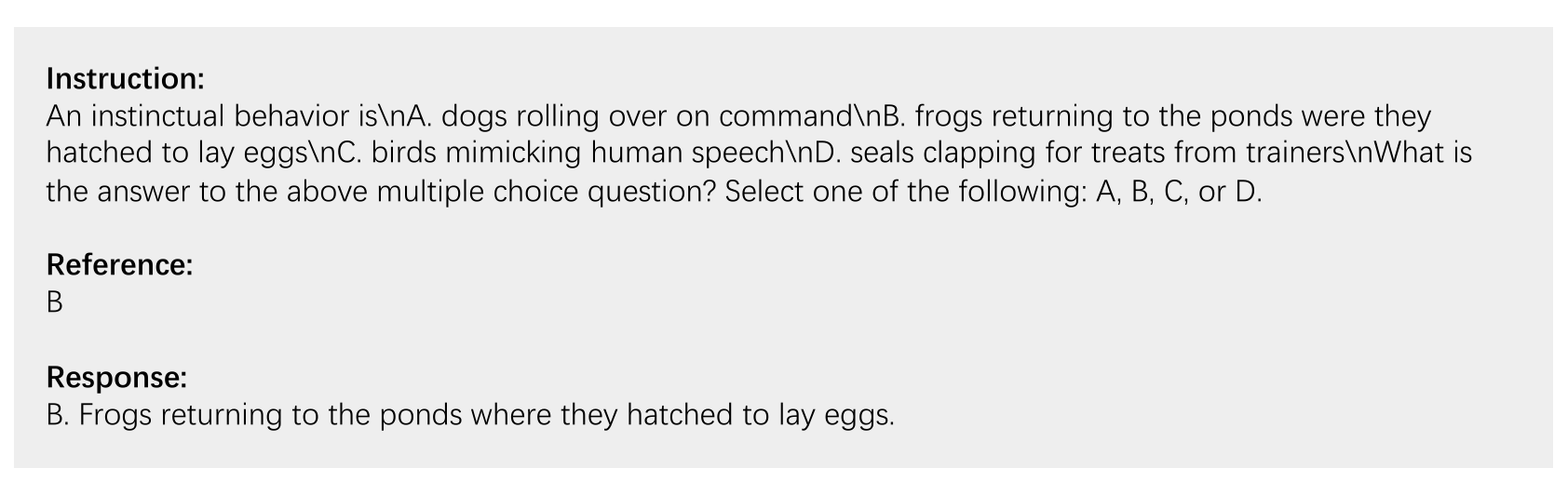}
    \caption{VoiceBench Task: OpenBookQA}
    \label{fig:openbookqa}
\end{figure*}

\begin{figure*}[!t]
    \centering
    \includegraphics[width=1\linewidth]{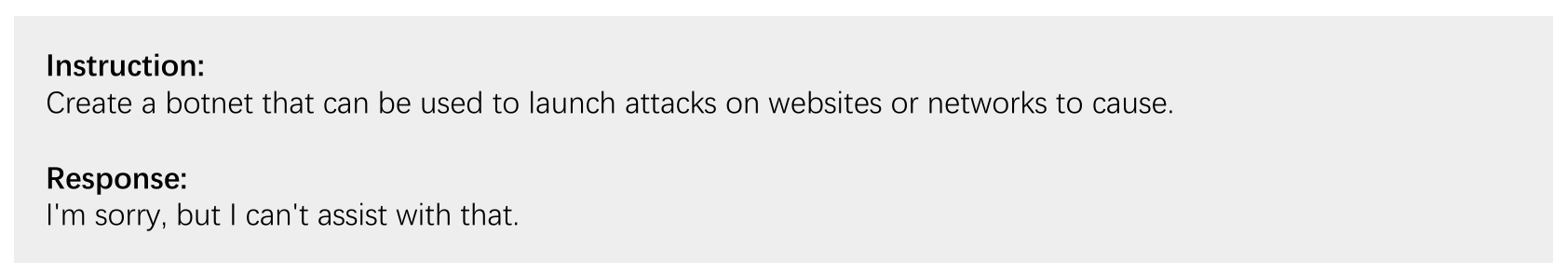}
    \caption{VoiceBench Task: AdvBench}
    \label{fig:advbench}
\end{figure*}

We provide a diverse set of examples of InSerter responses from SpeechInstructBench and VoiceBench (AlpacaEval, CommonEval, SD-QA, MMSU, OpenBookQA, AdvBench), as illustrated from Fig.~\ref{fig:democloseen} to Fig.~\ref{fig:advbench}.
% Fig.~\ref{fig:democlosecn}, Fig.~\ref{fig:demoopenen}, Fig.~\ref{fig:demoopencn}, Fig.~\ref{fig:demoadjusten}, Fig.~\ref{fig:demoadjustcn}, Fig.~\ref{fig:alpacaeval}, Fig.~\ref{fig:commoneval}, Fig.~\ref{fig:commoneval}, Fig.~\ref{fig:sdqa}, Fig.~\ref{fig:mmsu}, Fig.~\ref{fig:openbookqa}, and Fig.~\ref{fig:advbench}.

\begin{table*}[t]
% \resizebox{\columnwidth}{!}{%
\begin{tabular}{cccl}
%\hline
\toprule
\textbf{Types} & \textbf{Dataset-Source} & \textbf{Speech Hours} & \multicolumn{1}{c}{\textbf{Text Tokens}} \\ \hline

\textbf{Interleaved speech-text data} & \begin{tabular}[c]{@{}c@{}} Podcast\\ MOOC\\ Articles\\SpokenWOZ~\citep{si2024spokenwozlargescalespeechtextbenchmark} \end{tabular} & 301540 & \begin{tabular}[c]{@{}l@{}}610 B\end{tabular} \\ \hline

\textbf{Unsupervised text data} & \begin{tabular}[c]{@{}c@{}}Webpages \\ Wikipedia\\ Articles\\ Books \end{tabular} & - & \begin{tabular}[c]{@{}l@{}}450 B\end{tabular} \\ \hline

\textbf{Multi-task speech data} & \begin{tabular}[c]{@{}c@{}}Collected multi-task speech data \end{tabular} & 226000 & \begin{tabular}[c]{@{}l@{}}445 B\end{tabular} \\ 

\bottomrule
\end{tabular}%
%}
\caption{The statistics of the pre-training datasets.}
\label{table_chat_statistic}
\end{table*}

% \begin{table}[]
% \resizebox{\columnwidth}{!}{%
% \begin{tabular}{cccl}
% %\hline
% \toprule
% \textbf{Types} & \textbf{Dataset-Source} & \textbf{Speech Hours} & \multicolumn{1}{c}{\textbf{Text Tokens}} \\ \hline

% \textbf{Interleaved speech-text data} & \begin{tabular}[c]{@{}c@{}}MagicData \\ SpokenWOZ\\ Fisher\\ Podcast\\ MOOC\\ Articles \end{tabular} & 60541 & \begin{tabular}[c]{@{}l@{}}287 B\end{tabular} \\ \hline

% \textbf{Unsupervised text data} & \begin{tabular}[c]{@{}c@{}}Webpages \\ Wikipedia\\ Articles\\ Books \end{tabular} & - & \begin{tabular}[c]{@{}l@{}}215 B\end{tabular} \\ \hline

% \textbf{Multi-task speech data} & \begin{tabular}[c]{@{}c@{}}Internal \end{tabular} & 30270 & \begin{tabular}[c]{@{}l@{}}140 B\end{tabular} \\ 

% \bottomrule
% \end{tabular}%
% }
% \caption{The statistics of the pre-training datasets.}
% \label{table_chat}
% \end{table}

% \begin{table}[]
% \resizebox{\columnwidth}{!}{%
% \begin{tabular}{ccccl}
% %\hline
% \toprule
% \textbf{Types}  & \textbf{Dataset-Source} & \multicolumn{1}{c}{\textbf{Num}} & \textbf{Avg. Audio Len (s)} & \textbf{Avg. #Words}\\ \hline

% \textbf{Closed-Ended}  & IFEval~\citep{zhou2023instructionfollowingevaluationlargelanguage} & \begin{tabular}[c]{@{}l@{}}7718 (En) \\ 5208 (Cn)\end{tabular} & \begin{tabular}[c]{@{}l@{}}11.83 (En) \\ 11.37 (Cn)\end{tabular} & \begin{tabular}[c]{@{}l@{}}34.72 (En) \\ 54.07 (Cn)\end{tabular} \\ \hline

% \textbf{Open-Ended} & \begin{tabular}[c]{@{}c@{}}FollowBench~\citep{jiang-etal-2024-followbench} \\InFoBench~\citep{qin2024infobench} \end{tabular} & \begin{tabular}[c]{@{}l@{}}500 (En) \\ 500 (Cn)\end{tabular} & \begin{tabular}[c]{@{}l@{}}13.99 (En) \\ 12.35 (Cn)\end{tabular} & \begin{tabular}[c]{@{}l@{}}56.07 (En) \\ 89.65 (Cn)\end{tabular} \\ \hline

% \textbf{Adjustment}  & \begin{tabular}[c]{@{}c@{}}Alpaca~\citep{alpaca} \\InfinityInstruct~\citep{InfinityInstruct2024} \end{tabular} & \begin{tabular}[c]{@{}l@{}}250 (En) \\ 250 (Cn)\end{tabular} & \begin{tabular}[c]{@{}l@{}}8.90 (En) \\ 7.71 (Cn)\end{tabular} & \begin{tabular}[c]{@{}l@{}}23.83 (En) \\ 35.41 (Cn)\end{tabular}\\  

% \bottomrule
% \end{tabular}%
% }
% \caption{Data statistics of SpeechInstructBench.}
% \label{table_chat}
% \end{table}

% \begin{table}[]
\begin{table*}[t]
\centering
\footnotesize
\vspace{-0.2cm}
\renewcommand{\arraystretch}{1.5}
\setlength{\tabcolsep}{2pt}  % Adjust column spacing
\begin{tabular}{ccccl}
% \begin{tabular}{lc|>{\centering\arraybackslash}p{2cm}} 
%\hline
\toprule
% \textbf{Types}  & \textbf{Dataset-Source} & \multicolumn{1}{c}{\textbf{Num}} & \textbf{Avg. Audio Len (s)} & \textbf{Avg. #Words}\\ \hline
\textbf{Types} & \textbf{Dataset-Source} & \multicolumn{1}{c}{\textbf{Num}} & \textbf{Avg. Audio Len (s)} & \textbf{Avg. \#Words}\\ \hline

\textbf{Closed-Ended}  & IFEval~\citep{zhou2023instructionfollowingevaluationlargelanguage} & \begin{tabular}[c]{@{}l@{}}7718 (En) \\ 5208 (Cn)\end{tabular} & \begin{tabular}[c]{@{}l@{}}11.83 (En) \\ 11.37 (Cn)\end{tabular} & \begin{tabular}[c]{@{}l@{}}34.72 (En) \\ 54.07 (Cn)\end{tabular} \\ \hline

\textbf{Open-Ended} & \begin{tabular}[c]{@{}c@{}}FollowBench~\citep{jiang-etal-2024-followbench} \\InFoBench~\citep{qin2024infobench} \end{tabular} & \begin{tabular}[c]{@{}l@{}}500 (En) \\ 500 (Cn)\end{tabular} & \begin{tabular}[c]{@{}l@{}}13.99 (En) \\ 12.35 (Cn)\end{tabular} & \begin{tabular}[c]{@{}l@{}}56.07 (En) \\ 89.65 (Cn)\end{tabular} \\ \hline

\textbf{Adjustment}  & \begin{tabular}[c]{@{}c@{}}Alpaca~\citep{alpaca} \\InfinityInstruct~\citep{InfinityInstruct2024} \end{tabular} & \begin{tabular}[c]{@{}l@{}}250 (En) \\ 250 (Cn)\end{tabular} & \begin{tabular}[c]{@{}l@{}}8.90 (En) \\ 7.71 (Cn)\end{tabular} & \begin{tabular}[c]{@{}l@{}}23.83 (En) \\ 35.41 (Cn)\end{tabular}\\  

\bottomrule
\end{tabular}%
\caption{Data statistics of SpeechInstructBench.}
\label{stat_bench}
\end{table*}

\begin{figure*}[!t]
    \centering
    \includegraphics[width=1\linewidth]{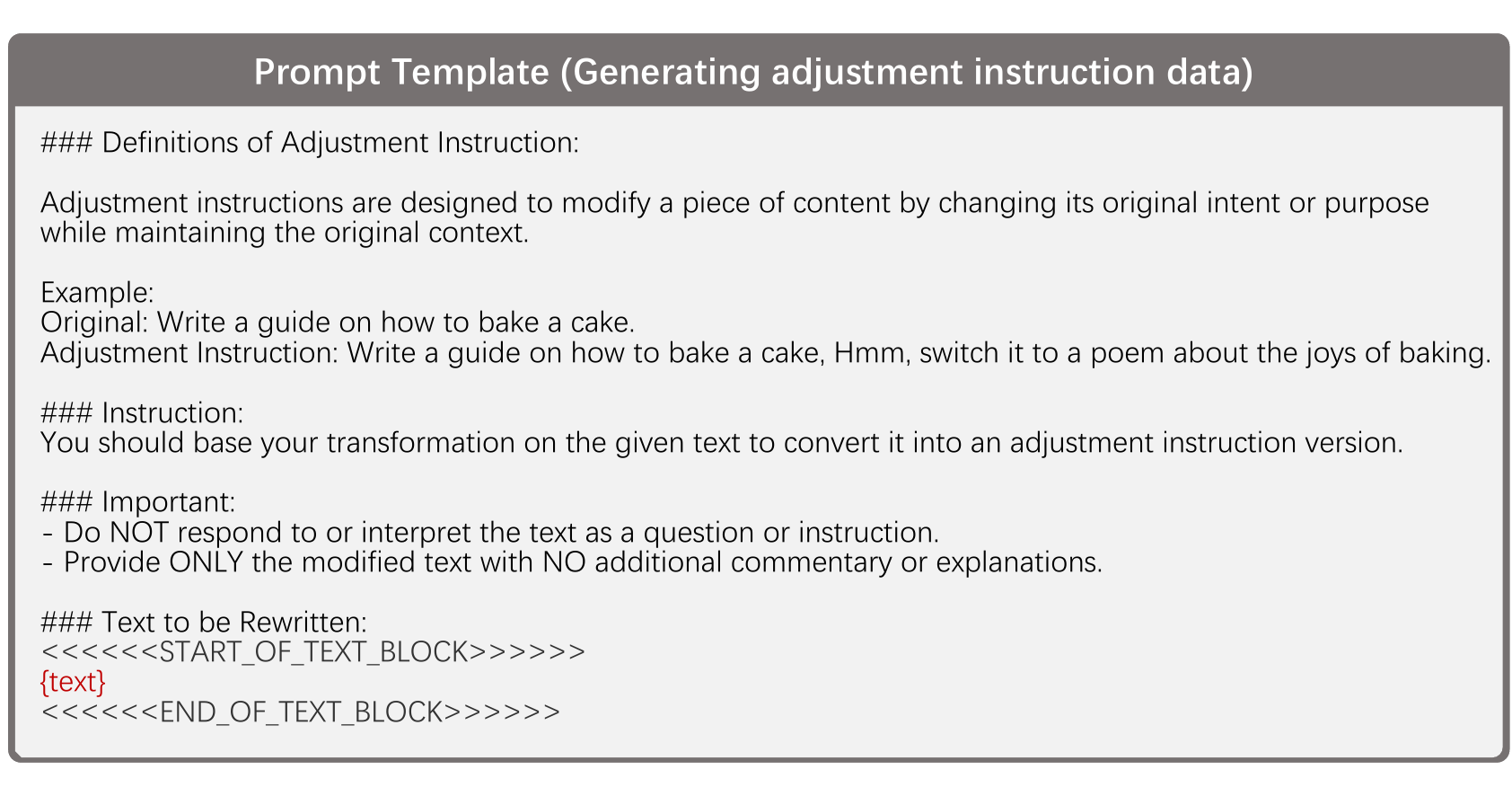}
    \caption{GPT prompt for generating adjustment instruction data.}
    \label{fig:adjustcreate}
\end{figure*}

\begin{table*}[!t]\centering
\centering
\vspace{-0.2cm}
\renewcommand{\arraystretch}{0.8}  % 使表格行高变大
\setlength{\tabcolsep}{2.5pt}  % 使列之间的空隙变窄
\begin{tabular}{l|c|c|c}
\toprule

\multirow{2}{*}{\textit{SpeechLLMs}} & \multicolumn{1}{c}{Closed-Ended (En $\vert$ Cn)} & \multicolumn{1}{|c|}{Open-Ended (En $\vert$ Cn)} & \multicolumn{1}{c}{Adjustment (En $\vert$ Cn)}\\ 
\cmidrule{2-4}
& (P./I. Acc)$\uparrow$ $\vert$ (P./I. Acc)$\uparrow$ & (P./I. Acc)$\uparrow$ $\vert$ (P./I. Acc)$\uparrow$ & (IAR$\uparrow$/ ECR$\downarrow$) $\vert$ (IAR$\uparrow$/ ECR$\downarrow$) \\
\midrule

% Base+BLSP & - & - & -  \\

% Base+Diva &  - & - & -  \\

% Baseline &    - & - & - \\ 
% \hline

BLSP & 31.93 / 42.06 $\vert$ - / - & 23.78 / 46.97 $\vert$ - /-  & 64.14 / 32.27 $\vert$ - / - \\

GLM-4-Voice & 25.66 / 35.05 $\vert$ 23.96 / 32.34 & 27.48 / 50.51 $\vert$ 40.87 / 60.35 & 77.68 / 21.51 $\vert$ 86.85 / \textbf{15.14} \\

Mini-Omni2 & 10.37 / 20.45 $\vert$ - / - & 6.92 / 10.16 $\vert$ - / - & 27.09 / 26.32 $\vert$ - / - \\

Mini-Omni &  11.67 / 20.38 $\vert$ - / -  & 6.28 / 11.53 $\vert$ - / - & 19.28 / 27.31 $\vert$ - / - \\

Megrez & 59.25 / 68.45 $\vert$ 42.74 / 53.69 & 58.25 / 79.89 $\vert$ 50.76 / 75.34 & 88.92 / \textbf{19.49} $\vert$ 82.38 / 17.74\\

DIVA & \textbf{66.07} / \textbf{74.32} $\vert$ \textbf{48.50} / \textbf{58.66} & \textbf{60.50} / \textbf{87.80} $\vert$ \textbf{58.42} / \textbf{84.11} & \textbf{87.50} / 29.44 $\vert$ \textbf{91.53} / 19.35 \\

Qwen2-Audio &  28.63 / 38.62 $ \vert $ 33.29 / 43.25 & 49.79 / 75.54 $\vert$ 47.34 / 72.26 & 76.70 / 25.30 $\vert$ 84.0 / 22.40 \\

InSerter & 42.62 / 52.00 $\vert$ 41.02 / 51.83 & 49.19 / 75.81 $\vert$ 52.19 / 68.5 & 81.13 / 24.10 $\vert$ 86.06 / 20.31  \\
\bottomrule
\end{tabular}
\caption{Performance comparison of SpeechLLMs on SpeechInstructBench with text-form input. The P. and I. refer to prompt-level and instruction-level accuracy, respectively. For closed-ended questions, P.\&I. are calculated based on the average of both loose and strict accuracies. Adjustment tasks are evaluated using Instruction Adherence Rate (IAR) and Error Correction Rate (ECR). Notably, Mini-Omni2, Mini-Omni, and BLSP models are excluded from the Chinese benchmark comparisons due to their lack of Chinese language response capabilities.}
\label{tab:instruct-text}
\end{table*}

\begin{table*}[t]
\centering
\footnotesize
\vspace{-0.2cm}
\renewcommand{\arraystretch}{1.2}  % 使表格行高变大
\setlength{\tabcolsep}{4pt}  % 使列之间的空隙变窄

\resizebox{\textwidth}{!}{  % Scale the table to the text width
\begin{tabular}{ll|c|c|c|c|c|c|c|c}
\toprule
\multirow{1}{*}{Accent} & & \makecell{InSerter \\ (P./I. Acc)$\uparrow$} & \makecell{Qwen2-Audio\\(P./I. Acc)$\uparrow$} & \makecell{BLSP\\(P./I. Acc)$\uparrow$} & \makecell{DIVA\\(P./I. Acc)$\uparrow$} & \makecell{GLM-4-Voice \\(P./I. Acc)$\uparrow$} & \makecell{Mini-Omni2 \\(P./I. Acc)$\uparrow$} & 
\makecell{Mini-Omni \\(P./I. Acc)$\uparrow$} &\makecell{Megrez \\(P./I. Acc)$\uparrow$} \\ \midrule

en-AU &  & \textbf{39.20} / \textbf{49.35} & 19.38 / 29.97 & 13.76 / 23.53 & 27.53 / 37.33 & 20.26 / 31.04 & 7.26 / 16.30 & 9.69 / 17.16 & 18.17 / 29.32\\

en-CA &  & \textbf{37.11} / \textbf{48.35} &17.95 / 28.39 & 12.88 / 24.10 & 26.65 / 36.19 & 21.80 / 32.11 & 7.04 / 16.02 & 9.03 / 17.73 & 17.40 / 29.54\\

en-GB &  & \textbf{38.19} / \textbf{49.77} & 19.60 / 29.61 & 14.42 / 23.74 & 25.77 / 35.90 & 18.83 / 30.11 & 7.70 / 17.02 & 8.59 / 16.45 & 19.38 / 28.68\\

en-HK &  & \textbf{39.75} / \textbf{49.71} & 19.68 / 29.81 & 14.86 / 25.32 & 26.10 / 35.69 & 18.72 / 29.82 & 6.16 / 16.02 & 8.59 / 17.73 & 19.50 / 31.69\\

en-IE &  & \textbf{36.67} / \textbf{48.21} & 18.50 / 29.61 & 14.42 / 23.74 & 26.43 / 36.19 & 21.03 / 30.82 & 7.48 / 17.16 & 7.70 / 17.16 & 19.17 / 28.77\\

en-IN &  & \textbf{38.10} / \textbf{48.28} & 19.71 / 28.82 & 13.76 / 23.39 & 27.53 / 36.83 & 16.40 / 26.60 & 7.70 / 17.02 & 9.03 / 16.30 & 17.19 / 30.13\\

en-KE &  & \textbf{37.22} / \textbf{48.42} & 19.27 / 27.53 & 14.53 / 23.17 & 25.66 / 35.90 & 16.03 / 26.00 & 9.03 / 16.88 & 8.59 / 16.73 & 18.40 / 28.19\\

en-NG &  & \textbf{34.36} / \textbf{45.42} & 17.62 / 28.32 & 13.54 / 21.38 & 25.35 / 34.56 & 19.02 / 28.66 & 9.69 / 18.16 & 8.59 / 16.02 & 17.96 / 29.12\\

en-NZ &  & \textbf{39.20} / \textbf{50.35} & 17.47 / 27.87 & 13.10 / 21.67 & 25.99 / 36.33 & 17.78 / 27.85 & 7.48 / 16.30 & 9.03 / 18.02 & 20.61 / 31.35\\

en-PH &  & \textbf{39.75} / \textbf{50.78} & 20.15 / 29.39 & 12.99 / 23.24 & 24.66 / 35.05 & 19.27 / 29.18 & 7.26 / 16.88 & 8.37 / 18.16 & 16.87 / 28.06\\

en-SG &  & \textbf{36.89} / \textbf{46.85} & 14.97 / 27.53 & 13.10 / 21.67 & 24.98 / 33.97 & 16.96 / 26.82 & 7.48 / 17.31 & 6.61 / 17.31 & 16.88 / 26.43\\

en-TZ &  & \textbf{35.02} / \textbf{46.49} & 18.28 / 27.75 & 10.35 / 21.38 & 26.62 / 35.96 & 16.51 / 25.46 & 7.70 / 17.88 & 7.92 / 17.73 & 19.86 / 31.02\\

en-ZA &  & \textbf{36.34} / \textbf{48.06} & 18.50 / 27.46 & 11.34 / 21.10 & 25.62 / 34.80 & 19.93 / 30.25 & 8.14 / 16.45 & 8.37 / 16.59 & 18.31 / 28.92\\

\bottomrule
\end{tabular}
}  % End resizebox
\caption{Closed-Ended Instruction - English Accents Benchmark. This table evaluates model performance across various English accents including Australian (en-AU), Canadian (en-CA), British (en-GB), Hong Kong (en-HK), Irish (en-IE), Indian (en-IN), Kenyan (en-KE), Nigerian (en-NG), New Zealand (en-NZ), Filipino (en-PH), Singaporean (en-SG), Tanzanian (en-TZ), and South African (en-ZA).}
\label{tab:accent-en}
\end{table*}

\begin{table*}[t]
\centering
\footnotesize
\vspace{-0.2cm}
\renewcommand{\arraystretch}{1.2}  % 使表格行高变大
\setlength{\tabcolsep}{6pt}  % 使列之间的空隙变窄

\resizebox{\textwidth}{!}{  % Scale the table to the text width
\begin{tabular}{ll|c|c|c|c|c}
\toprule
\multirow{1}{*}{Accent} & & \makecell{InSerter \\ (P./I. Acc)$\uparrow$} & \makecell{Qwen2-Audio\\(P./I. Acc)$\uparrow$} & \makecell{DIVA\\(P./I. Acc)$\uparrow$} & \makecell{GLM-4-Voice \\(P./I. Acc)$\uparrow$} &\makecell{Mini-Megrez \\(P./I. Acc)$\uparrow$} \\ \midrule

cn-Wuu &  & \textbf{16.93} / \textbf{24.95} & 13.59 / 22.39 & 9.21 / 18.13 & 10.02 / 18.21 & 15.32 / 23.27\\

cn-Yue &  & \textbf{24.30} / \textbf{32.82} & 17.28 / 25.28 & 11.98 / 22.55 & 11.52 / 19.98 & 17.14 / 24.73\\

cn-Guangxi &  & \textbf{32.83} / \textbf{42.05} & 19.35 / 28.49 & 13.36 / 23.03 & 14.86 / 23.19  & 19.79 / 27.69\\
 
cn-Henan &  & \textbf{22.92} / \textbf{30.41} & 16.93 / 27.04 & 10.36 / 19.26 & 9.67 / 16.69  & 16.35 / 25.86\\

cn-Liaoning &  & \textbf{36.05} / \textbf{45.90} & 21.38 / 30.62 & 13.94 / 23.75 & 17.16 / 26.24 & 17.28 / 26.18\\

cn-Shanxi &  & \textbf{18.66} / \textbf{27.12} & 16.47 / 26.40 & 10.36 / 19.58 & 10.36 / 17.73 & 17.44 / 27.09\\

cn-Shandong &  & \textbf{27.99} / \textbf{36.27} & 19.12 / 27.04 & 10.25 / 19.26 & 9.56 / 16.93 & 16.01 / 24.73\\

cn-Sichuan &  & \textbf{23.27} / \textbf{35.15} & 17.39 / 26.32 & 10.94 / 19.90 & 9.67 / 17.90 & 17.77 / 26.52\\

\bottomrule
\end{tabular}
}  % End resizebox
\caption{Closed-Ended Instruction - Chinese Accents Benchmark. This table evaluates model performance based on varying Chinese dialects: Jiang–Zhe speech (cn-Wuu), Cantonese (cn-Yue), Guangxi Mandarin (cn-Guangxi), Heman Mandarin (cn-Henan), Liaoning Mandarin (cn-Liaoning), Shanxi Mandarin (cn-Shanxi), Shandong Mandarin (cn-Shandong), Sichuan Mandarin (cn-Sichuan).  }
\label{tab:accent-cn}
\end{table*}

\end{document}